\documentclass[a4paper,fleqn,usenatbib]{mnras}

\usepackage{mathptmx}

\usepackage[T1]{fontenc}
\usepackage{ae,aecompl}

\usepackage{graphicx}	
\usepackage{amsmath}	
\usepackage{amssymb}	
\usepackage{pdflscape}

\title[Millimeter line observations toward local galaxies]{Millimeter line observations toward four local galaxies}

\author[Fei Li et al.]{Fei Li,$^{1,2,3}$\thanks{E-mail: lifei@shao.ac.cn}
Junzhi Wang,$^{1,4}$\thanks{E-mail: jzwang@shao.ac.cn}
Minzhi Kong,$^{3}$\thanks{E-mail: confucious\_76@163.com}
Shanghuo Li,$^{1,2,5}$
\\
$^{1}$Shanghai Astronomical Observatory, Chinese Academy of Sciences, 80 Nandan Road, Shanghai 200030, China \\
$^{2}$University of Chinese Academy of Sciences, 19A Yuquanlu, Beijing 100049, China \\
$^{3}$Hebei Normal University, No.20 Road East. 2nd Ring South, Yuhua District, Shijiazhuang, Hebei, 050024, China\\
$^{4}$Key Laboratory of Radio Astronomy, Chinese Academy of Sciences,   Nanjing, 210008,  China\\
$^{5}$Harvard-Smithsonian Center for Astrophysics, 60 Garden Street, Cambridge, MA 02138
}

\date{Accepted XXX. Received YYY; in original form ZZZ}

\pubyear{2017}

\begin{document}
\label{firstpage}
\pagerange{\pageref{firstpage}--\pageref{lastpage}}
\maketitle

\begin{abstract}

We present results of millimeter line observations toward four local gas-rich galaxies (NGC 3079, NGC 4258, NGC 6240 and VII Zw 31) with the IRAM 30 meter millimeter telescope. More than 33 lines in these four sources were detected, including normal dense gas tracers (HCN 1-0, HCO$^+$ 1-0, and C$_2$H 1-0, etc) and their isotopic species. H$^{13}$CN (1-0) and H$^{13}$CO$^+$ (1-0) are detected for the first time in NGC 4258. Optical depths of HCN 1-0 and HCO$^{+}$ 1-0 were estimated with detected  isotopic lines in NGC 4258, which were 4.1 and 2.6, respectively. 
 HC$_3$N $J=29-28$, which requires high volume density and high temperature to excite, was detected in NGC 6240. High ratios of HCO$^+$/HCN in NGC 4258 and NGC 6240 imply that this ratio  might  not be  a perfect diagnostic tool between AGN and starburst environments, due to contamination/combination of both processes. The low HC$_3$N/HCN line ratios with less than 0.15 in NGC 4258, NGC 6240 and the non-detection of HC$_3$N line in NGC 3079 and VII Zw 31 indicates that these four galaxies are HC$_3$N-poor galaxies. The variation of fractional abundance of CN in different types of galaxies is large.
\end{abstract}


\section{Introduction}

Molecular interstellar medium (ISM) plays an essential role in star formation \citep{2007RPPh...70.1099O}, where  star formation takes place \citep{1979IAUS...84...35S}. Molecular gas is not only as fuel in the active process of galaxies but should also, in turn, be strongly affected by the activity \citep{2008ApJ...677..262K}. Molecular line observations, based on millimeter facilities, are crucial for understanding the physical and astro-chemical conditions of galaxies, which can provide plenty of information about density, mass, temperature, kinematics, magnetic field and so on \citep{1998ARA&A..36..317V}.

Multi-line studies of  local galaxies have been made in literature  \citep{2017ApJ...835..265W,2015A&A...573A.116M,2013MNRAS.436..570D,2013A&A...549A..39A,2004ApJ...606..271G}.  Observational and theoretical works \citep{2009ApJ...692.1432G,2009ApJ...694..610M}  showed that the thermal and astro-chemical structures of gas in nearby galaxies should differ between starburst (SB) and active galactic nucleus (AGN) dominated regions. The activity in galaxy can be either an starburst or an AGN, or a combination of both. Gas properties derived from molecular lines can infer which physical process of activity  dominates the heating of  molecular cloud. Some studies had been done to distinguish between AGNs and starburst signatures of centers of galaxies \citep{2001ASPC..249..672K,2004AJ....128.2037I,2008ApJ...677..262K}.  Because the existence of both starburst and AGN can create exceptional and extreme conditions for the molecular gas in the central region, more spectral lines should be detected in AGN+SB composite galaxies than pure AGN-dominated galaxy. However, the connection between AGN and starburst are still not well understand. Local galaxies with both AGN and SB activities, are ideal targets for multiline studies of heavy rotor molecules. 

In this paper, we present results of millimeter spectroscopy observations toward  two nearby galaxies: NGC  3079, NGC 4258,  and two Ultra-luminous infrared galaxies (ULIRGs): NGC 6240, VII Zw 31. The outline of this paper is as follows: we introduce the properties of the four sources in \S2 the observations and data reduction in \S3,  and describe the main results of the data in \S4. In \S5, we present the analysis and discussion of the results. And a brief summary is provided \S6.

\section{Source selection}

We select four local galaxies to study molecular gas properties, with different nature of nuclear activities.  NGC 3079 and NGC 4258 are two nearby type II AGN with strong H$_2$O maser emission, while  NGC 6240 and VII Zw 31 are two gas-rich (U)LIRGs. Our original proposals were searching for SiO mega-masers in  NGC 3079 and NGC 4258 at 3 mm band from about 85 GHz to 92 GHz,  and searching for red-shifted 118.75 GHz molecular oxygen in  NGC 6240 and VII Zw 31, which covers the rest frequencies from about 111 GHz to 119 GHz. NGC 3079, NGC 4258 and NGC 6240 are well studied and have relatively bright molecular emission lines.

NGC 3079, classified as a type II Seyfert galaxy \citep{2001ApJS..136...61S},  is an edge-on spiral galaxy with optical and radio nuclear activity at the distance of 16.1 Mpc \citep{2015yCat..51500043T}. The central region not only hosts a AGN, but also hosts a powerful starburst. Dense gas tracers (CN, HCO$^+$, and HNC 3-2 \& 1-0 lines) of this source were reported by \cite{2007A&A...476..177P},  which concluded that emissions of HNC and HCN emerge from the same gas and the ratios of HCN/HNC and CN/HCN favor a PDR scenario, rather than an XDR.  HCO$^+$/HCN 1-0 line ratio is 1.12, while  HNC/HCN 1-0 line ratio is 0.27 \citep{2011A&A...528A..30C}. Molecule lines of C$_2$H, CS, C$^{18}$O, $^{13}$CO, CO at J = 1-0 transition and CN 1-0,  as well as   marginal detections of  CH$_3$OH 5-4 and HC$_3$N 12-11 were reported in \cite{2011A&A...528A..30C} and \cite{2011AJ....141...38S}.

NGC 4258 (M106) is a Sab-type LINER/Seyfert galaxy at the distance of 6.8 Mpc \citep{2015yCat..51500043T}. Its anomalous spiral arms emerge from the galaxy nucleus and appear to intersect the regular spiral arms of galaxy \citep{2014ApJ...788L..33O}. \cite{2007A&A...467.1037K} reported that NGC4258 is a weak AGN and a pair of radio jets may be influencing the molecular emission. HCO$^+$ 1-0, H41$\alpha$ and $^{13}$CO 1-0 lines were reported in \cite{2011AJ....141...38S}, while \cite{2011MNRAS.418.1753J} reported  the detections of HCO, HCN, C$_2$H, HCO$^+$ 1-0 and tentatively detected H$^{13}CO^+$ 1-0 at $ < 2 \sigma$ level.  They both showed a very large HCO$^{+}$/HCN ratio in NGC 4258.

NGC 6240 is a local ULIRG with extreme starburst activity and prominent AGN \citep{2003ApJ...582L..15K} at the distance of 98 Mpc \citep{2012A&A...538A.152Z}, as merger of two AGN host galaxies. Observations showed that molecular emissions are concentrated in a small region between the two Seyfert nuclei \citep{2008ApJS..178..189W,2007ApJ...659..283I}. Dense gas tracers including lines of HCN, HCO$^+$, HNC and other molecular lines of CO, C$_2$H, HNCO, C$^{18}$O and CN 1-0 had been detected, as well as tentatively detection of H$^{13}$CN, SiO, H$^{13}$CN and CH$_3$OH lines and HC$_3$N 10-9 and 12-11 were reported in \cite{2009ApJ...692.1432G}. \cite{2008ApJ...677..262K} and \cite{2011A&A...528A..30C} showed that the ratio of HCO$^{+}$/HCN 1-0 was about 1.6 and suggested that HCO$^{+}$/HCN is not a reliable tracer of XDRs.


VII Zw 31 is a gas-rich ULIRG \citep{1986MNRAS.219P...1F} at the distance of 240 Mpc \citep{2011ApJ...739L..25L}. It is a possible protogalactic disk at a low redshift and a merger-induced starburst \citep{1990AJ.....99.1414D}. This source had one of the highest known CO luminosities and its molecule gas may be a large part of the dynamical mass \citep{1987ApJ...321L.103S,1991ApJ...368L..15R}. Detections of dense gas tracers of HCN 1-0 \citep{2004ApJS..152...63G}  and CS 5-4  \citep{Wang11} were also reported in literature. \cite{1998ApJ...507..615D} presented  detections of CO 1-0 and 2-1 and tentatively detections of CN 1-0.



\section{Observations and data reduction}

Our observations toward  four local galaxies were made with the IRAM 30 meter millimeter telescope at Pico Veleta, Spain\footnote{Based on observations carried out with the IRAM 30m Telescope. IRAM is supported by INSU/CNRS (France), MPG (Germany) and IGN (Spain).}, in December 2012. The setup of the observations were shown in Table  \ref{tabel 1}.
The Eight Mixer Receiver (EMIR) with dual-polarization and the Fourier Transform Spectrometers (FTS) backend with  8 GHz frequency coverage and 195 kHz frequency spacing were used. Standard wobbler switching mode with a $\pm$120$''$ offset at 0.5 Hz beam throwing was used for the observations.  Pointing  was checked about every 2 hours with nearby strong millimeter emitting quasi-stellar objects. 
The EMIR receiver can provide two bands simultaneously. During the observations for each source with different days, we used  3 mm band with similar frequency tuning to cover mega-masers or red-shifted O$_2$ emission,  while different frequency tuning setups were used at 1mm band to cover lines we were interested. The observational parameters are summarized in Table \ref{tabel 1}, with typical system temperature  of about 137 K at 3 mm band and 142 K at 1mm band, including about 40 K from the receiver.  About 12 hours observing time (on + off) were spent for each source of  NGC 4258 and NGC 3079.  Since  they are the extremely nearby  objects, the beam ($\sim$29$''$ at 3mm) of IRAM 30 meter  only covered the  central region of NGC 4258 and NGC 3079. On the other hand,  the observing time (on+off) toward  VII Zw 31 and NGC 6240 is about 11 hours for each source. Emissions of  our observations are from the entire galaxy for both VII Zw 31 and NGC 6240, as they are  distant enough to be smaller than the beam of  IRAM 30 meter telescope.

All the data were reduced with the CLASS package of GILDAS\footnote{http://www.iram.fr/IRAMFR/GILDAS}. We checked the line profile visually, and qualified spectra by comparing the measured noise and the theoretical noise before and after a few times of boxcar smooth. About 5\% of the spectra were discarded during the qualification. We averaged them with time weighting and subtracted linear baselines for all spectra.  We converted the antenna temperature (T$_{A}^{*}$) to the main beam brightness temperature (T$_{mb}$), using T$_{mb}$=T$_{A}^{*}\cdot$F$_{eff}$/B$_{eff}$, where F$_{eff}$ and B$_{eff}$ from the IRAM 30m homepage \footnote{http://www.iram.es/IRAMES/mainWiki/Iram30mEfficiencies}  is tabulated in Table \ref{tabel 2}.

\begin{table}
\begin {center} 
\caption{Telescope parameters from the IRAM 30m homepage$^{~\rm 3}$}	\label{tabel 2}
\begin{tabular}{p{3cm}p{1.5cm}p{1.5cm}}
\hline
\hline
{Observed Frequency} & {F$_{eff}$} & {B$_{eff}$} \\         
\hline
\hline
85GHz	      &	0.95	 &	0.81\\	
111GHz	&	0.94	 &	0.78\\
220GHz	&	0.92	 &	0.59\\
217GHz	&	0.94	 &	0.63\\
229GHz	&	0.92      &	0.59\\	
248GHz	&	0.92      &	0.59\\
255GHz	&	0.87      &	0.49\\			          
\hline	  
\end{tabular}
\end{center}
\end{table}

\begin{table*}

   \centering
   \begin{minipage}{160mm}
\caption{Observational  parameters}	
      \label{tabel 1}
		\begin{tabular}{r r r r r r r r}
   	         \hline
	         \hline
	         {Source} & {RA} & {DEC} & {Band range} & {Tsys}&{Time}& {RMS}&{$\delta_v$}\\
                      &	(J2000)  &   (J2000) & (GHz) & (K) & (min) & (mK) & (km\,s$^{-1}$) \\
	         \hline
	          \hline
NGC3079	&	10:01:57.80	 &	55:40:47.0	& 85-92      & 114    & 115    &  4     & 0.7\\
	    &                &               &  220-227   &  260   & 131    & 13     & 0.3\\
NGC4258 &	12:18:57.50	 &	47:18:14.0   & 85-92      &  101   &  650   &  2     & 0.7\\
        &                 &               &  219-227   &  210   &  102   & 8      & 0.3\\
        &                 &               &  229-236   &  151   &  122   & 10     & 0.3\\
NGC6240 &	16:52:58.90 	  &	02:24:03.0	 &  111-120	  & 167    &  175   & 6      & 0.2\\
        &                 &               &  260-268	  & 232    &  100   & 15     & 0.5\\
VIIZW31 &	05:16:45.10	  &	79:40:13.0 	 & 112-117    &  117    & 89    &5        & 0.5\\
	    &                 &               & 228-236    &  292   & 145    & 13      & 0.3\\
	    &                 &               & 260-268    &  198   & 110    & 11      & 0.2\\ 

	         \hline	  
		\end{tabular}\\	
		
Notes. 
The broad-band observations include 1 mm and 3 mm band. $\delta_v$ is the velocity resolution of the molecular lines relative to 195 kHz frequency resolution, while the 'RMS' in each line was obtained with 'base' at this velocity resolution. 
\end{minipage}
 \end{table*}

\begin{table*}
\centering
  \begin{minipage}{160mm}
\caption{Line ratios} 
\label{tabel 3}
\begin{tabular}{c c c c c c c c c c c}
\hline
\hline
{Sources} & {$\dfrac{I_{1-0}^{HCO^+}}{I_{1-0}^{HNC}} $} & {$\dfrac{I_{1-0}^{HCO^+}}{I_{1-0}^{HCN}} $} & {$\dfrac{I_{1-0}^{HNC}}{I_{1-0}^{HCN}} $} & {$\dfrac{I_{3-2}^{HCO^+}}{I_{3-2}^{HCN}} $} & {$\dfrac{I_{29-28}^{HC_3N}}{I_{3-2}^{HCN}} $} & {$\dfrac{I_{10-9}^{HC_3N}}{I_{1-0}^{HCN}} $} \\
\hline
\hline
NGC 3079  &	 $3.17 \pm 0.51$	 &	$0.96 \pm 0.11$  &  $0.30 \pm 0.04$ &  . . .  & . . . & $ < 0.03 $\\	
NGC 4258  &	 $6.88	\pm 0.69$  &	$1.99 \pm 0.06$    &  $0.29 \pm 0.03$ &  . . .  & . . . &$ 0.11 \pm 0.03$ \\
NGC 6240  &	. . .	       &	. . .    &  . . .    & $1.48 \pm 0.08$  & $0.03 \pm 0.01$ & . . .\\
VII Zw 31	&	. . .	       &	. . .    &  . . .    & $0.78 \pm 0.15$  &$<0.2$& . . .\\
		          
\hline	
Literature Data\\
\hline
{Sources} & {$\dfrac{I_{3-2}^{HCO^+}}{I_{3-2}^{HCN}} $} &{$\dfrac{I_{1-0}^{HCO^+}}{I_{1-0}^{HCN}} $} & {$\dfrac{I_{1-0}^{HNC}}{I_{1-0}^{HCN}} $}  &  {$\dfrac{I_{10-9}^{HC_3N}}{I_{1-0}^{HCN}} $} & {Type}&{Reference}\\
NGC 1068 & $0.38\pm0.07 $  &$0.60 \pm 0.01$ & . . .                    &. . .& AGN&1 \\
NGC 4388 &  . . .   &$1.38\pm 0.4 $   &$0.62\pm0.24$  & $<0.35$ &AGN&2\\
NGC 5194 &$<0.7$ &$0.72\pm0.06$   &. . .                     &. . .          &AGN&1\\
NGC 7469 &. . .&$1.12\pm 0.11$   &$0.55\pm 0.08$ & $<0.07$ &AGN&2\\
NGC 2273 &. . .&$1.05\pm0.37$    &$1.09\pm0.38$  & $<0.89$ &AGN&2\\
NGC 4826 &$<0.7$&$0.59\pm0.04$    &. . .                      &. . .          &AGN+SB&1\\
NGC 3627 & $<0.7$&$1.0\pm0.1$        &. . .                      &. . .          &AGN+SB&1\\
NGC 4569 &$<0.9$  &$0.85\pm0.07$    &. . .                      &. . .          &AGN+SB&1\\
NGC 6951 &$<0.8$  &$0.7\pm0.05$      &. . .                      &. . .          &AGN+SB&1\\
NGC 6946 & $0.94\pm0.08$  &$0.86\pm0.07$    &. . .                      &. . .          &SB&1\\
NGC 2146 &$1.2\pm0.2$   &$1.3\pm0.08$      &. . .                      &. . .          &SB&1\\
M 82(center)& $0.7\pm0.2 $&$1.5\pm0.04$    &. . .                      &. . .          &SB&1\\
NGC 1614    &. . .&$1.83\pm0.37$   & $0.33\pm0.15$   &$ <0.35$ & SB&2\\
NGC 4194   &  . . . &$1.32\pm0.32$    &$0.53\pm0.23$    &$<0.20$  & SB&2\\
NGC 660   &  . . .  &$1.04\pm0.09$     &$0.52\pm0.06$    &$<0.07$  & SB&2\\
NGC 3556& . . . &$1.57\pm0.37$     &$0.37\pm0.24$    &$<0.27$  & SB&2\\
NGC 6240 & $0.7\pm0.2$   &$1.6\pm0.2$         &. . .                      &. . .          &ULIRG(AGN+SB)&1\\
Mrk 231  &$0.4\pm0.1$&$0.87\pm0.09$     &. . .                      &. . .          &ULIRG(AGN+SB)&1\\
Arp 220   &$0.2\pm0.05$&$0.48\pm0.05$     &. . .                      &. . .          &ULIRG(AGN+SB)&1\\
NGC 4418 &. . . &$0.59\pm0.1 $      &$0.47\pm0.09$   &$<0.18$  &LIRG(AGN)&2\\
		          
\hline

\end{tabular}\\
1 Data taken from \cite{2008ApJ...677..262K}. \\2 Data taken from \cite{2011A&A...528A..30C}. \\Our result shown that the ratio of HCO${+}$/HCN (J = 3-2) in NGC 6240 is different from \cite{2008ApJ...677..262K} reported.
\end{minipage}
\end{table*}

\section{Results}

\subsection{Detected Lines}

A total of 33 molecular transitions were detected in these four galaxies with this observation.
Two isotopic lines, H$^{13}$CN 1-0 and H$^{13}$CO$^{+}$ 1-0,  were detected for the first time in NGC 4258, while these two lines, as isotopic lines of dense gas tracers,  were not detected in NGC 3079, which can only give the  upper limits (see Table \ref{Table:NGC3079}).  HC$_3$N  was firstly detected in NGC 6240 with $J$=29-28 transition, while  HNCO was firstly detected in VII Zw 31 with 12-11 transition. The observational parameters of each line were summarized in Table \ref{Table:NGC3079}, \ref{Table:NGC4258}, \ref{Table:NGC6240}, and \ref{Table:VII Zw 31}, including  central velocity,  line widths,  peak intensity, velocity integrated  intensity and column density. Lines with velocity integrated flux above 3 $\sigma$ level  are considered as  detections. The line identification was made using the Lovas \citep{1992JPCRD..21..181L} catalogues, and the Splatalogue data base\footnote{www.splatalogue.net}, which is a compilation of line lists including the Jet Propulsion Laboratory \citep{1998JQSRT..60..883P}. Informations of each source were  described  below. 


For NGC 3079, as a composite source with both AGN and starburst, there are  plenty of molecular lines in this source (See Figure \ref{fig:NGC3079} and Table \ref{Table:NGC3079}). Normal dense gas tracers  (HCO$^{+}$ 1-0, HCN 1-0, HNC 1-0,  C$_2$H 1-0, and two groups of CN 2-1) were detected, as well as   HNCO 4-3, CH$_3$CN 12$_{0}$-11$_{0}$  as new detections. Except for CH$_3$CN and $^{18}$CO, other all species present a double peck structure (See Figure \ref{fig:NGC3079} ), which are consistent with that in the literature \citep{2007A&A...476..177P}.

c-C$_3$H$_2 $ 2-1, as well as  isotopic lines of dense gas tracers, i.e.,  H$^{13}$CN 1-0, H$^{13}$CO$^{+}$ 1-0 (See Figure \ref{fig:NGC4258} and Table \ref{Table:NGC4258}) were new detections in NGC 4258. However, CN 2-1 and C$^{18}$O lines were not detected.



\begin{figure*}
	\centering
\includegraphics[width=5in]{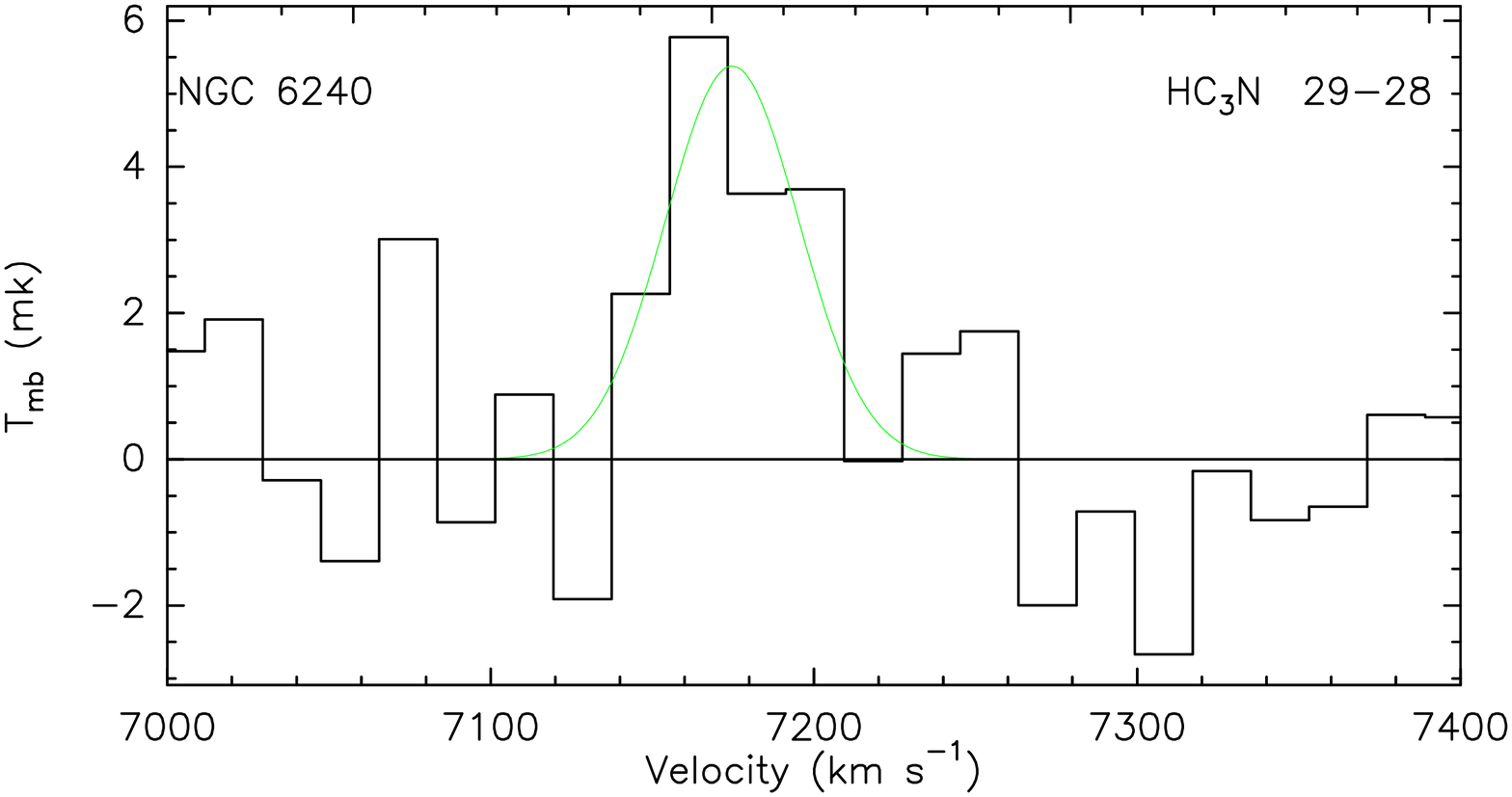}

\vspace*{-0.2 cm} \caption{HC$_3$N 29-28 line detected in NGC 6240, with rms of 1.07 mK at the velocity resolution of 23.97 km\,s$^{-1}$, overlaid with Gaussian fitting (green line).
	\label{fig:NGC6240_isotope}}
    \vskip-10pt
\end{figure*}

HC$_3$N (29-28), which is optically thin and requires a dense and warm cloud component to excite, was firstly detected in NGC 6240 and showed a narrow line width of about 48 km s$^{-1}$ (See Figure \ref{fig:NGC6240_isotope}). Both red-shifted and blue-shifted line wings of CO 1-0 were detected in NGC 6240, which indicate possible molecular outflow from AGN in this source, which was consistent with the results in literature \citep{2013A&A...558A..87F}. For VII Zw 31, HNCO 12$_{1,11}$-11$_{0,10}$  (see Figure \ref{fig:VIIZW31} and Table \ref{Table:VII Zw 31}) and  C$_2$H 3-2 (see Figure \ref{fig:NGC6240_C2H,VIIZW31_C2H}) were detected as new results. While other molecular lines had been reported in  literature for these two sources (See  Table \ref{Table:NGC6240},\ref{Table:VII Zw 31}).



We compared the intensity ratio of HCO$^+$/HCN with the previous observation in these four sources, which suggested that the ratio of HCO$^+$/HCN might not be an ideal tool to identify AGN or starburst in galaxies.

\begin{figure*}
	\centering
\includegraphics[width=5in]{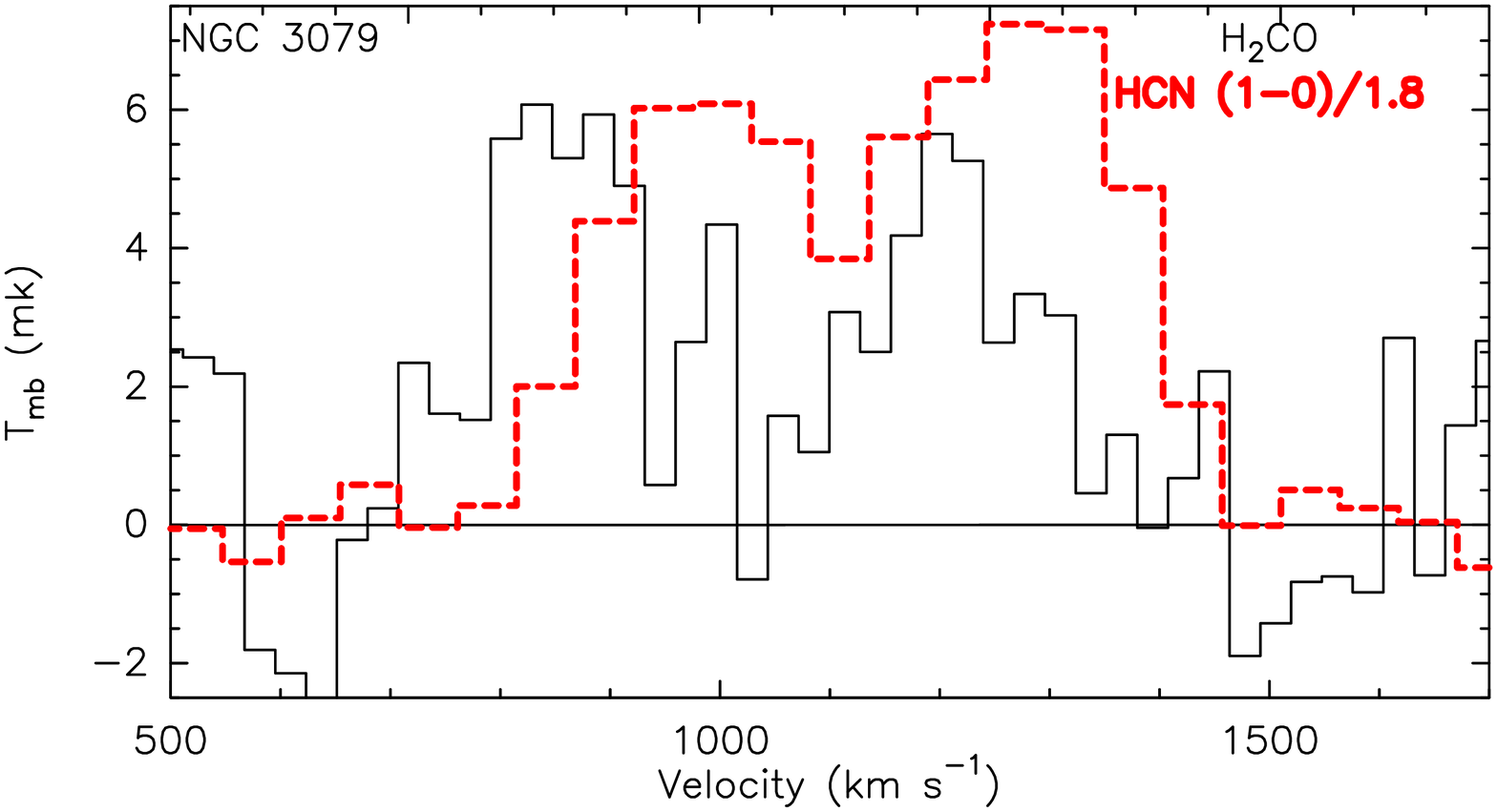}

\vspace*{-0.2 cm} \caption{Two emission features near 225.7 GHz with identification as H$_2$CO $(3_{1,2}-2_{1,1})$ (black line), overlaid with HCN (1-0) divided by 1.8 (red line).  
	\label{fig:NGC3079_H2CO}}
    \vskip-10pt
\end{figure*}

\subsection{Column densities}

Local thermodynamic equilibrium (LTE)  and optically thin emission for all the detected transitions were assumed to  calculate the molecular column densities ($N_{mol}$) using the Boltzmann diagrams method \citep{1999ApJ...517..209G}, which needs at least two transitions of a given molecule to calculate the rotational temperatures. However,  only one transition of each molecule was detected in our data for the four sources. A fixed $T_{ex}$  was used for each source: 30 K for NGC 3079, 50 K for NGC 4258, 49 K for NGC 6240 \citep{1997A&A...326..963B,1996ApJ...473..219C,2007ApJ...659..283I}, and 49 K for VII Zw 31,  which  was adopted the same excitation temperature of NGC 6240  and was higher than the brightness temperature of 13 K from observation of CO \citep{1991ApJ...368L..15R}, were used to calculate the column densities.
The data also were corrected by the beam dilution effect, as $T_{B}$=[($\theta_{s}^2$+$\theta_{b}^2)$/$\theta_{s}^2]\times T_{MB}$, where $T_{B}$ is the source average brightness temperature, $\theta_{s}$ is the source size, $\theta_{b}$ is the beam size in arc seconds, and $T_{MB}$ is the measured main beam temperature. Based on the four sources interferometric observations of CO (1-0) for NGC 3079 \citep{2002ApJ...573..105K}, CO (2-1) NGC 4258 \citep{2007ApJ...658..851S}, CO (1-0) for NGC 6240 \citep{2013A&A...558A..87F}, CO (1-0), CO (2-1) for VII Zw 31 \citep{1998ApJ...507..615D}, we roughly assumed the sizes of $11''$ and $8''$ for the nuclear regions of  NGC 3079 and NGC 4258;  $3''$ and $4''$ for  NGC 6240 and VII Zw 31, respectively.

At LTE and optically thin conditions, we have 
 \begin{equation}
\label{eq:pop}
N_u=\frac{8\pi k \nu^2 W}{h c^3 A_{u\ell}}
\end{equation}
where $N_u$ is the upper state column density of the molecule, $W$ is the integrated brightness temperature, $k$ is the Boltamann constant, $\nu$ is the frequency of the transition, h is the  Planck constant and $A_{u\ell}$ is the Eintein's transition probability coefficients. The spectroscopic parameters were obtained from the Cologne Database for Molecular Spectroscopy (CDMS; \cite{2005JMoSt.742..215M} ) and JPL catalogs \citep{1998JQSRT..60..883P}. 
\begin{equation}
\label{eq:lte}
N_u=\frac{N}{Q(T_{ex})}g_ue^{-E_u/kT_{rot}}
\end{equation}
where $N$ is the total number density of molecules, $Q(T_{ex} $) is the partition function, $g_u$ is the degeneracy of the upper state and $E_u$ is the upper level energy. The partition function $Q(T_{ex} $) of each molecule is calculated by fitting the partition function at different temperature from CDMS.

The column densities for all molecules were showed in Tables \ref{Table:NGC3079}, \ref{Table:NGC4258}, \ref{Table:NGC6240}, and \ref{Table:VII Zw 31}. Note that although we assumed all lines are optically thin for calculation, some lines are certainly optical thick, such as HCN and HCO$^{+}$ in NGC 3079 and NGC 4258 (See Table \ref{tabel 4}), for which the column densities were underestimated. Ratios of column densities for the four sources were calculated and listed in Table \ref{tabel 9}.

\begin{figure*}
	\centering
\includegraphics[width=5in]{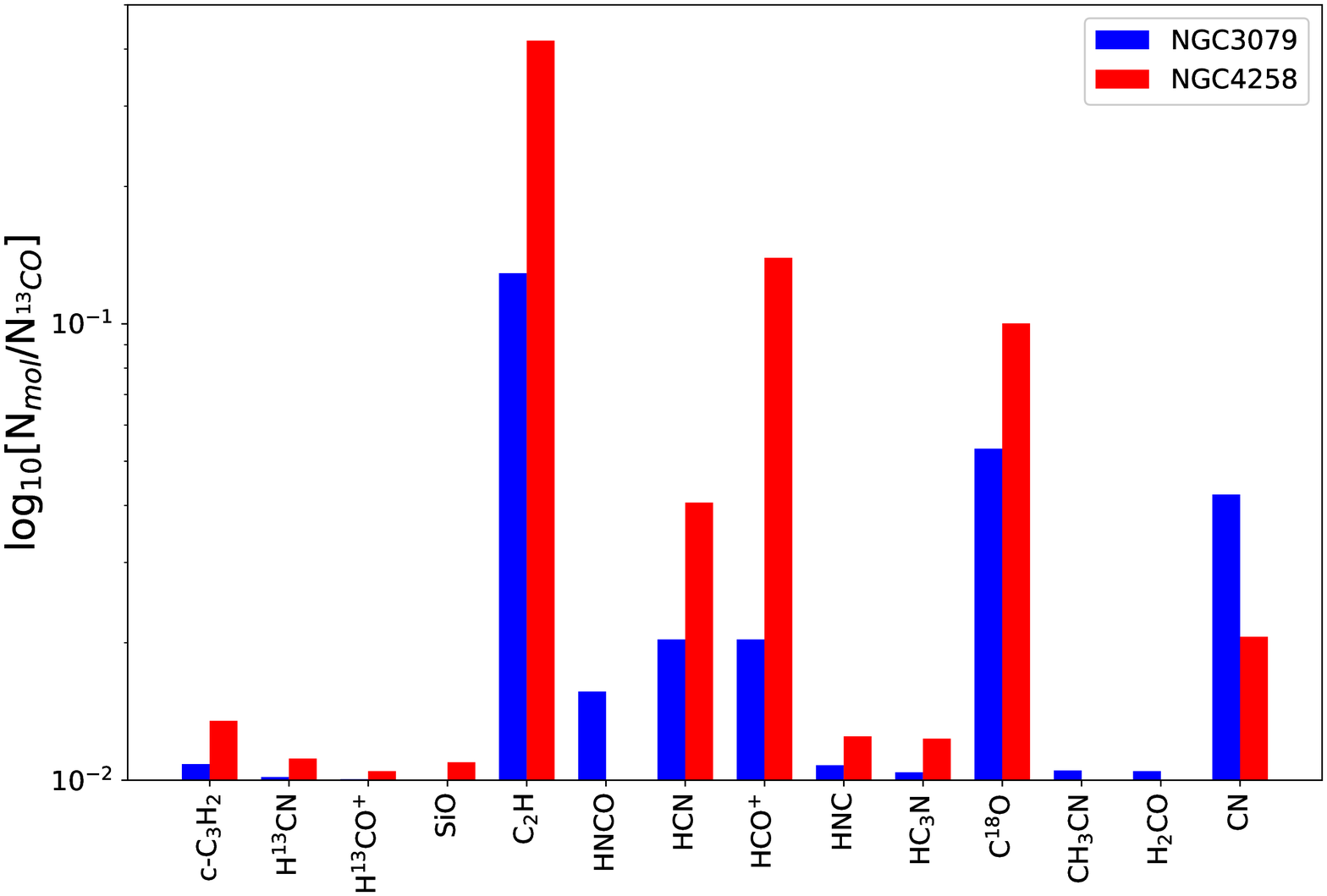}

\vspace*{-0.2 cm} \caption{Comparison between NGC 3079 and NGC 4258 fractional abundances of different molecules with respect to the $^{13}$CO abundances. For $^{18}$CO, the lower limit just be shown in NGC 4258, while  CO is not shown here because it is off the chart.
	\label{fig:colum_3079-4258}}
    \vskip-10pt
\end{figure*}

\begin{figure*}
	\centering
\includegraphics[width=5in]{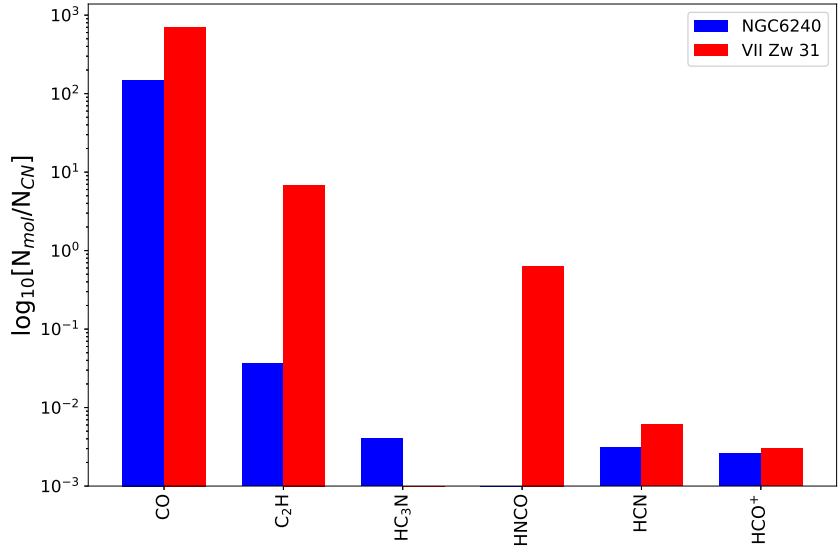}

\vspace*{-0.2 cm} \caption{Comparison between  NGC 6240 and VII Zw 31 fractional abundances of different molecules with respect to the CN abundances. 
	\label{fig:colum_6240-zw31}}
    \vskip-10pt
\end{figure*}

\section{Discussions}

\subsection{NGC 3079 v.s. NGC 4258: AGN and starburst hybrid v.s. pure AGN}
\label{sec:AGN}

As mentioned in \S3, the IRAM beams can not cover the entire galaxy for both NGC 3079 and NGC 4258, which cause that the  detected emissions  are mainly from the nuclear region.


\subsubsection{Molecules}

The comparison of molecular abundances among these two galaxies was shown in Figure \ref{fig:colum_3079-4258}. HNCO $4_{0,4}-3_{0,3}$ at 87.925 GHz,  CH$_3$CN  $ 12_{0}-11_{0} $ at 220.747 GHz, H$_2$CO  $ 3_{1,2}-2_{1,1}$ at 225.698 GHz, and the two CN 2-1 lines at 226.66 GHz and  226.875 GHz, were detected only in NGC 3079.Starburst active in NGC 3079 may enhanced the abundances of these species. Since NGC 3079 is more distant than NGC 4258, the intrinsic luminosities of these line emissions are significantly higher than the upper limits in NGC 4258. Shock gas tracer and dense gas tracer are detected in NGC 3079, such as HNCO and CH$_3$CN, as well as PDR tracer H$_2$CO. The relative abundances of C$_2$H, HNCO, HCN, HCO$^+$ and HNC  to CN are similar to that in  NGC 253 \citep{2015A&A...579A.101A}, which implies that the environment of starburst in NGC 3079 might be similar to that in NGC 253. NGC 3079 may be in an intermediate starburst stage, with an active starburst. Shock gas tracer HNCO had higher abundances than PDR tracers c-C$_3$H$_2$ in NGC 3079. All above species had also been detected toward nearby starburst galaxy  M82 \citep{2011A&A...535A..84A} and the nearest ULIRGs Arp 220 \citep{2011A&A...527A..36M}.



On the other hand, the faint lines of  H$^{13}$CN 1-0, H$^{13}$CO$^+$ 1-0, and SiO 2-1, were only detected in NGC 4258. The velocity integrated intensity and line center of SiO 2-1 of NGC 4258 is similar to that  of  \citet{2013ApJ...778L..39W}, but with broader line width. SiO usually trace shocks in extragalactic sources \citep{2000A&A...355..499G,2001ApJ...563L..27G} and its abundance may be enhanced by X-ray chemistry in AGNs \citep{2013A&A...549A..39A}. That is, SiO is somehow expected to be detected in NGC 3079, which have strong active AGN and shock gas. However, it was not detected at similar noise level of NGC 4258. All of the lines  detected in NGC 3079 and NGC 4258 had been detected as strong emission lines in Orion A with line survey \citep{1985ApJS...58..341S}. It is interesting that CN 2-1 in NGC 4258 as a pure AGN was not detected, which was contrary to models showed by \cite{2007A&A...461..793M}. While the strong emission of CN 2-1 was detected in NGC 3079 (SB+AGN), in which CN might be enhanced not only in XDRs, but also PDRs near massive young stars \citep{2002A&A...381..783A}. C$_2$H 1-0 emissions were also detected in NGC 4258 and  NGC 3079. However, the 6 hyperfine structures can not be separated due to the line broadening  (See Figure \ref{fig:NGC3079_C2H,NGC4258_C2H}).

\begin{figure*}
	\centering
\includegraphics[width=3in]{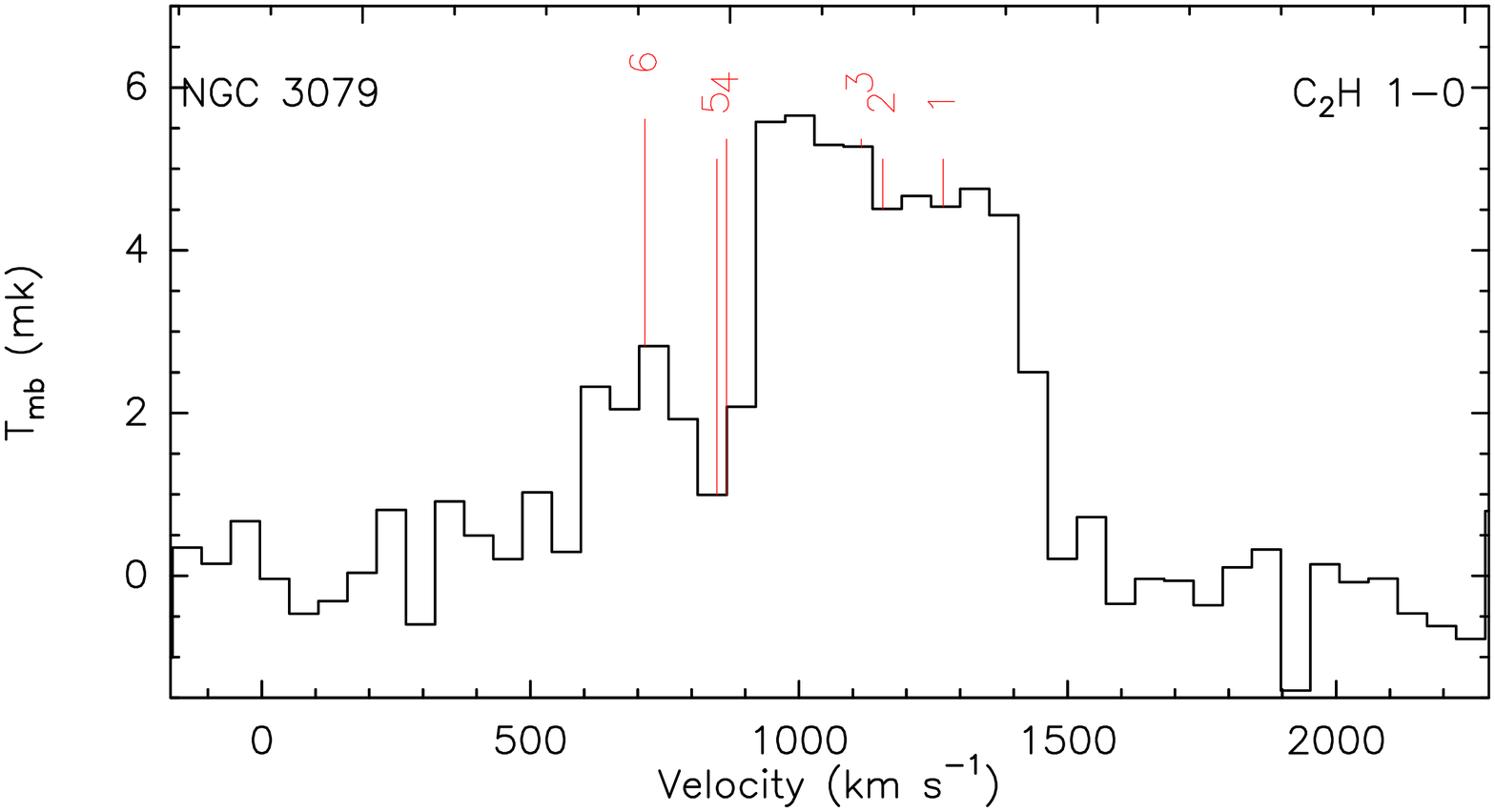}
\includegraphics[width=3in]{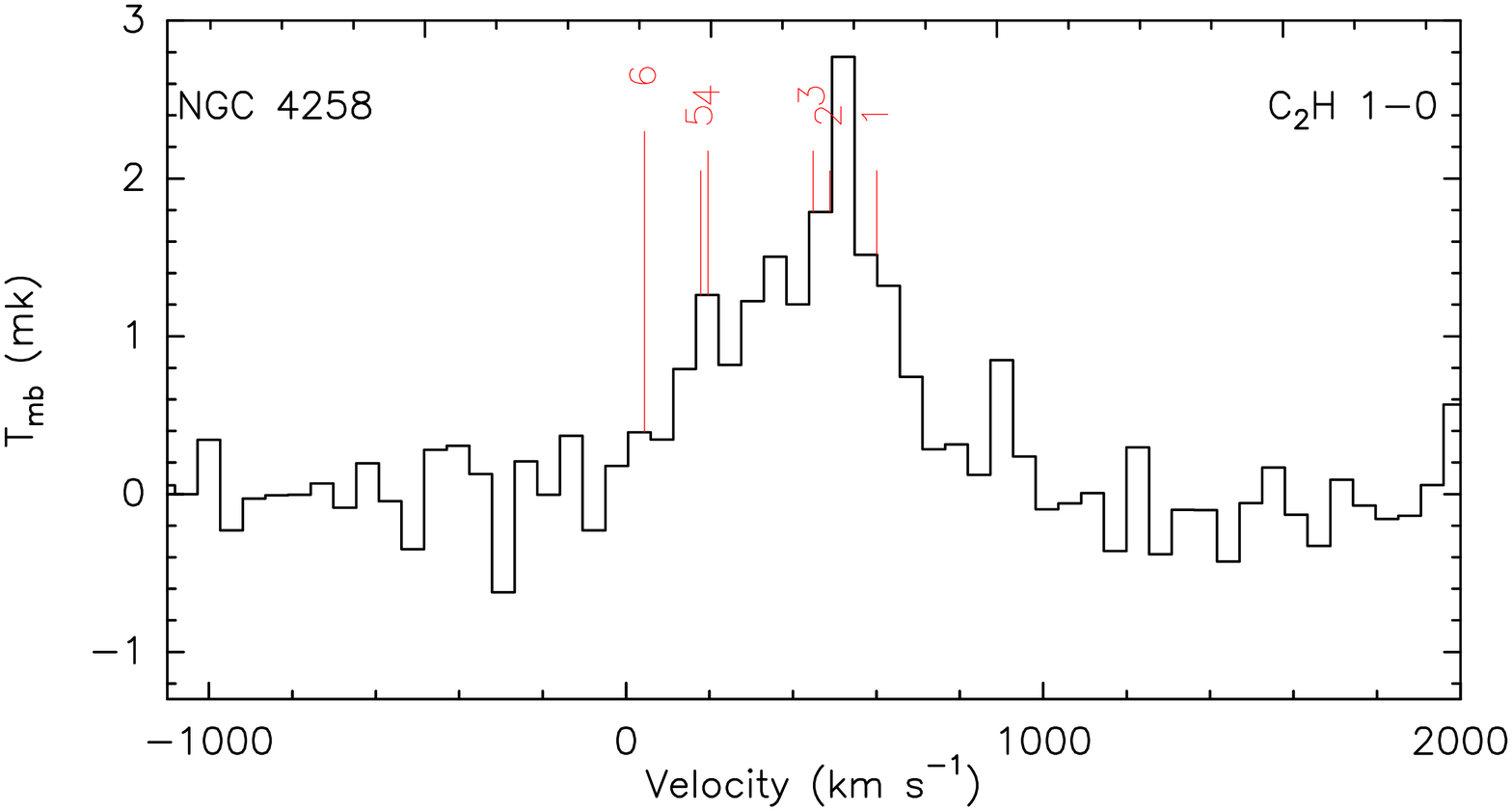}

\vspace*{-0.2 cm} \caption{C$_2$H (1-0) detected in NGC 3079 and NGC 4258. The rms is 0.55 mK at the velocity resolution of 54.31 km\,s$^{-1}$ in NGC 3079 and 0.25 mK at the velocity resolution of 54.31 km\,s$^{-1}$ in NGC 4258. The same six hyperfine lines are marked from 1-6 in these two sources: 1, C$_2$H 1-0 3/2-1/2 F=1-1 at 87284.15GHz; 2, C$_2$H 1-0 3/2-1/2 F=2-1 at 87316.92GHz; 3, C$_2$H 1-0 3/2-1/2 F=1-0 at 87328.62GHz; 4, C$_2$H 1-0 1/2-1/2 F=1-1 at 87402.00GHz; 5, C$_2$H 1-0 1/2-1/2 F=0-1 at 87407.16GHz; 6, C$_2$H 1-0 1/2-1/2 F=1-0 at 874446.47GHz. The transition of J=1-0 3/2-1/2 F=1-0 is used as reference for the velocities in these two galaxies. 
	\label{fig:NGC3079_C2H,NGC4258_C2H}}
    \vskip-10pt
\end{figure*}

\begin{figure*}
	\centering
\includegraphics[width=3in]{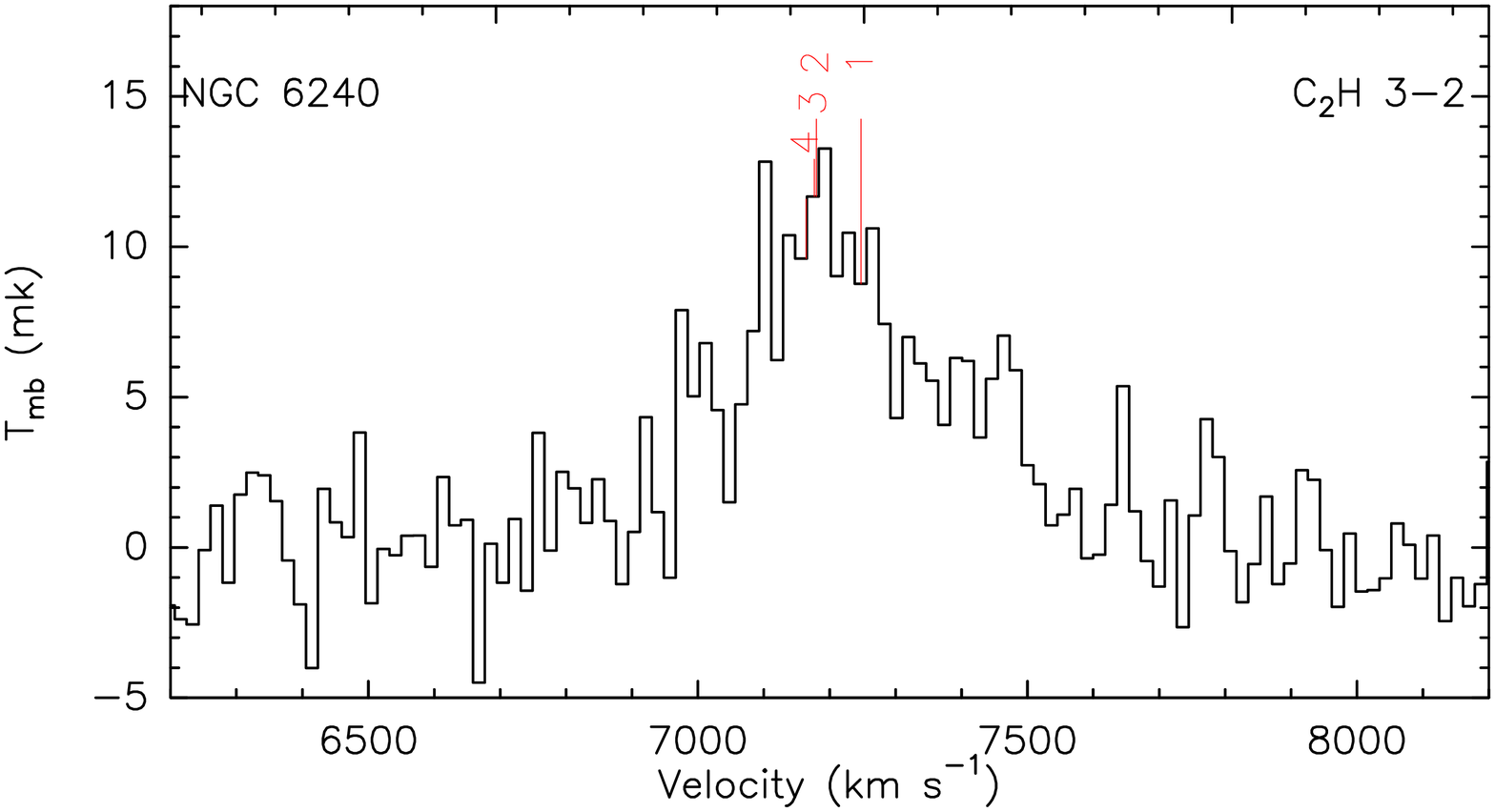}
\includegraphics[width=3in]{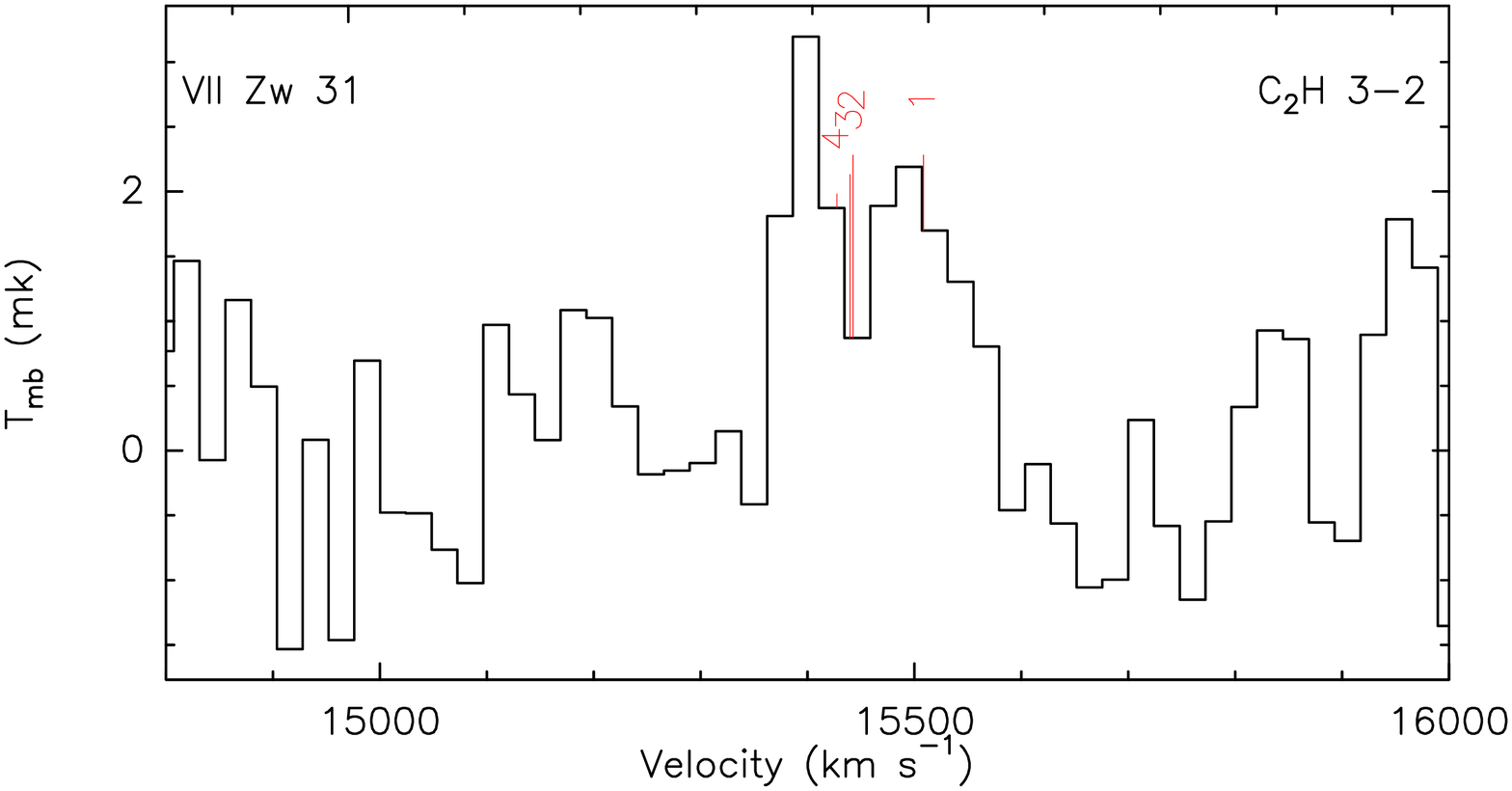}

\vspace*{-0.2 cm} \caption{C$_2$H (3-2) detected in NGC 6240 and VII ZW 31. The  rms is 1.89m K at the velocity resolution of 18.1 km\,s$^{-1}$ in NGC 6240 and 1.18 mK at the velocity resolution of 18.10 km\,s$^{-1}$. The same four hyperfine lines are marked from 1-4 in these two sources: 1, C$_2$H 3-2 J=7/2-5/2 F=4-3 at 262.004 GHz; 2, C$_2$H  3-2 J=5/2-3/2 F=3-2 at 262.065 GHz; 3. C$_2$H 3-2 J=5/2-3/2 F=2-1 at 262.067 GHz, 4, C$_2$H 3-2 J=5/2-3/2 F=2-2 at 262.079 GHz.  The transition of J=5/2-3/2 F=2-1 is used as reference for the velocities in these two galaxies.
	\label{fig:NGC6240_C2H,VIIZW31_C2H}}
    \vskip-10pt
\end{figure*}




\subsubsection{The line profiles}
\label{sec:width}

\begin{figure*}
	\centering 
\includegraphics[width=2.8in]{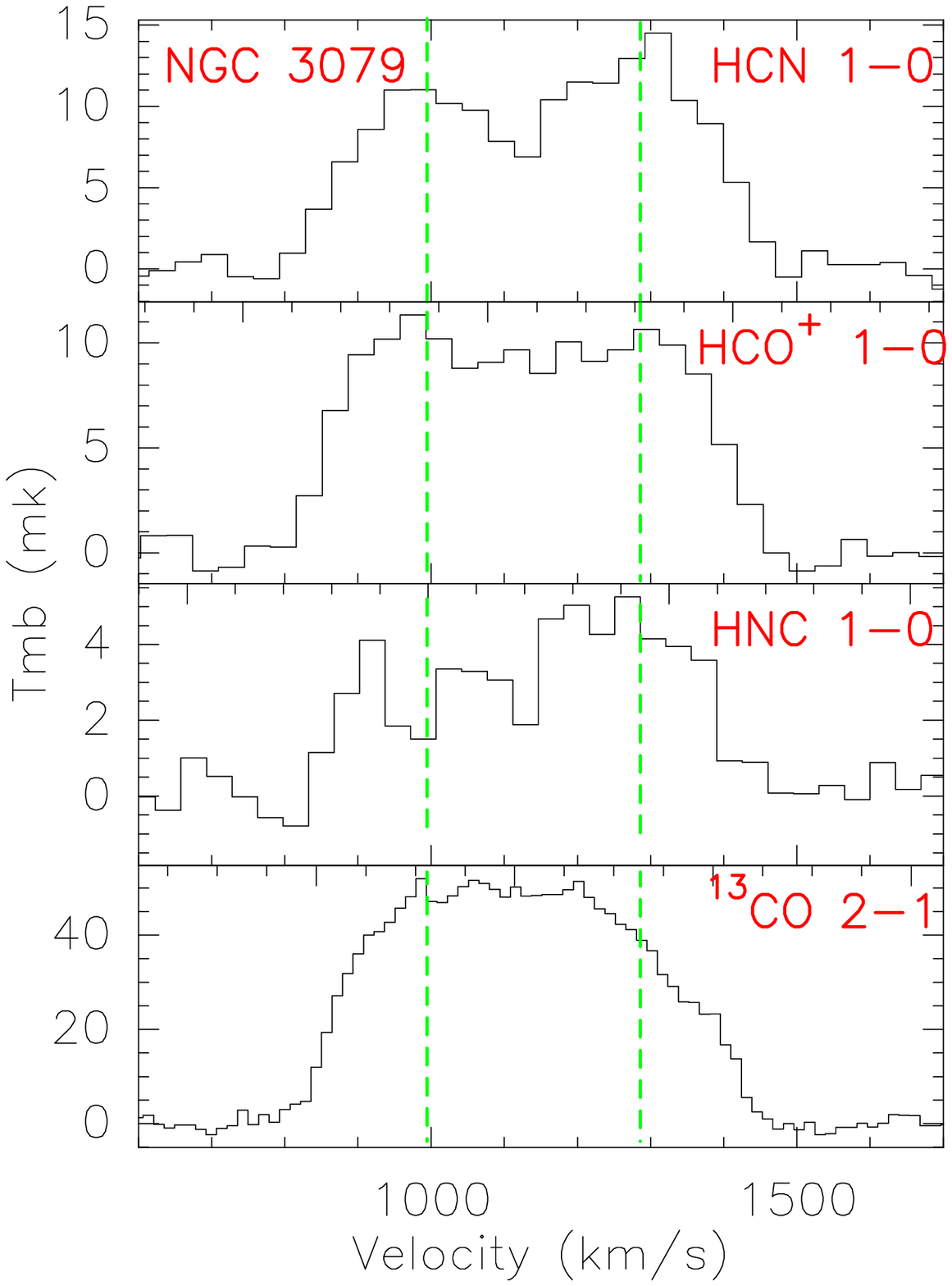}
\includegraphics[width=2.9in]{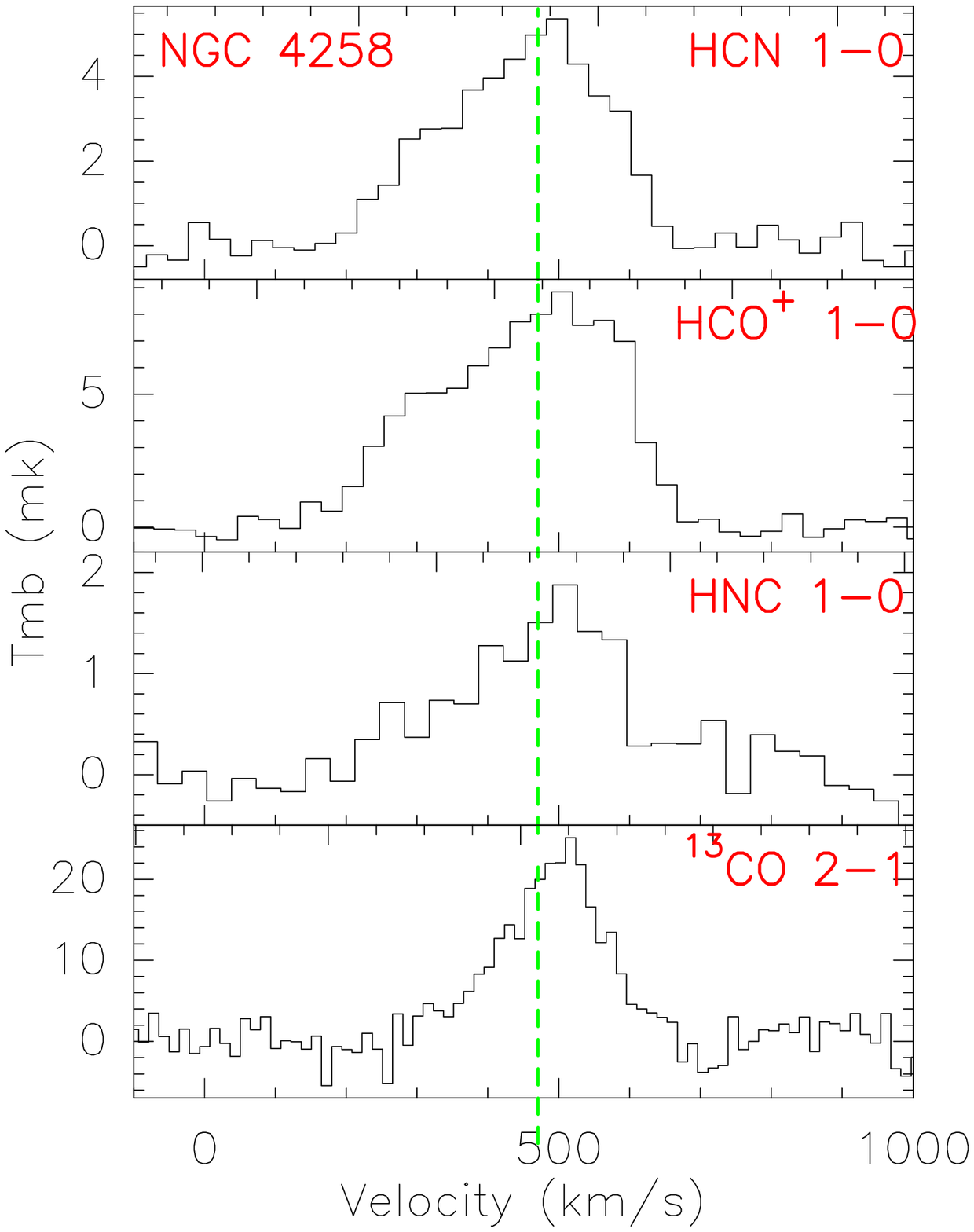}

\vspace*{-0.2 cm} \caption{Spectra lines in NGC 3079 and NGC 4258.
	\label{fig:spec}}
    \vskip-10pt
\end{figure*}

The line parameters for NGC 3079 and NGC 4258, including velocity-integrated flux, line velocity and line width, are summarized in Table \ref{Table:NGC3079}, \ref{Table:NGC4258}. For NGC 3079, except for CH$_3$CN and $^{18}$CO, other all species have showed obviously two separated velocity components, which is consistent with the results in literature \citep{2007A&A...476..177P} (See Figure \ref{fig:spec}). It indicates that two  components with different velocities exist in this source. $^{18}$CO 2-1 may be a double peak structure. However, it is at the edge of the band, which causes only the blue-shifted component was detected, while the red-shifted component was out of the frequency coverage. Based on profiles of other lines, such as HCN 1-0 and HCO$^+$ 1-0, $^{13}$CO 2-1 should have two components velocity. However, it is hard to distinguish the double peak structure, due to the high opacity. CH$_3$CN could have two velocity components. However, we can not distinguish CH$_3$CN emission profile due to the red-shifted component of CH$_3$CN is blended by $^{13}$CO 2-1. On account of most of red-shift component of molecular lines were detected in NGC 3079, we used line widths of  red-component of HCN 1-0 to estimate upper limit of  integrated intensities for HC$_3$N 10-9 and  H$^{13}$CN 1-0, while use  that of  HCO$^+$ 1-0 for H$^{13}$CO$^+$ 1-0.



For NGC 4258, we find that except for H$^{13}$CN, H$^{13}$CO$^+$, H$_{3}$CN and SiO, the central velocity of other species is about 450 km/s and the line width is about 250 km/s. H$^{13}$CO$^+$ and H$^{13}$CN are just detected around 3$\sigma$ level, which may cause the uncertainty of estimating velocity. 

 

\subsubsection{Line ratios}
\label{sec:ratio}

The ratios of HCO$^{+}$/HCN, HCO$^{+}$/HNC, and HCN/HNC were listed in Table \ref{tabel 3}. HCO$^+$/HCN line ratio had been used to distinguish between AGNs and starburst signatures in galactic centers \citep{2001ASPC..249..672K,2007AJ....134.2366I,2008ApJ...677..262K}. As a result in  literature, HCO$^{+}$/HCN in starburst and LIRGs was higher than that in AGN, explained by  the enhancement of HCN abundance in XDR surrounding an AGN \citep{2011A&A...528A..30C,2017ApJ...835..213P}. Furthermore, an enhanced intensity of HCN (1-0), with respect to HCO$^+$ (1-0) in the studies of AGN dominated galaxies, was interpreted as evidence for the influence of X-ray dominated regions (XDRs) or mechanical heating \citep{2016ApJ...818...42I}. But there are some counterarguments to this statement that lower line ratio of HCO$^{+}$/HCN, which were reported  in pure starburst and composite galaxies \citep{2011A&A...528A..30C,2015ApJ...814...39P}. \cite{2013ApJ...779...47M} presented  that low HCO$^{+}$/HCN ratio was also found in individual galactic star-forming regions. Although many different effects can contribute to an HCN enhancement in active environments, the actual cause is not clear.


\cite{2013PASJ...65..100I,2016ApJ...818...42I} concluded that the high abundances and mechanical heating from a jet or shock could drive the HCN enhancement relative to HCO$^+$ in AGN galaxies. While, in our observations, the high line ratio of HCO$^{+}$/HCN 1-0 ($\sim$2) in NGC 4258,  which is consistent with that in \cite{2011MNRAS.418.1753J}, is different to the typical value of pure AGNs \citep{2008ApJ...677..262K}. Different optical depth of these two molecules might play a key role in this ratio. The  opacity of  HCN was almost twice of HCO$^+$ (see Table \ref{tabel 4}). On the other hand, the lower critical densities of HCO$^+$ than that of HCN, may cause that HCO$^+$ region from the relative low density gas. This scenario is equivalent to that discussed toward local starbursts galaxies for the high HCO$^{+}$/HCN ratio \citep{2012ApJ...755..104M}.  The ratio of HCO$^{+}$/HCN in NGC 3079 was consistent with that in \cite{2011A&A...528A..30C}, similar to that in  other AGN/starburst composite sources in literature \citep{2008ApJ...677..262K}.

Since the red-shifted component of C$^{18}$O 2-1 was out of frequency coverage, we just use peak temperature to estimate the ratio of $^{13}$CO/ C$^{18}$O as 3.5. On the other hand, based on the non-detection of C$^{18}$O 2-1, the lower limit of this ratio in NGC 4258 is about 11. The different line ratios of $^{13}$CO/C$^{18}$O in these two galaxies might due to different isotopic ratios of $^{13}$C/$^{12}$C and  $^{16}$O/$^{18}$O, or self-absorption of $^{13}$CO 2-1 in NGC 3079.

\begin{figure*}
	\centering 
\includegraphics[width=3in]{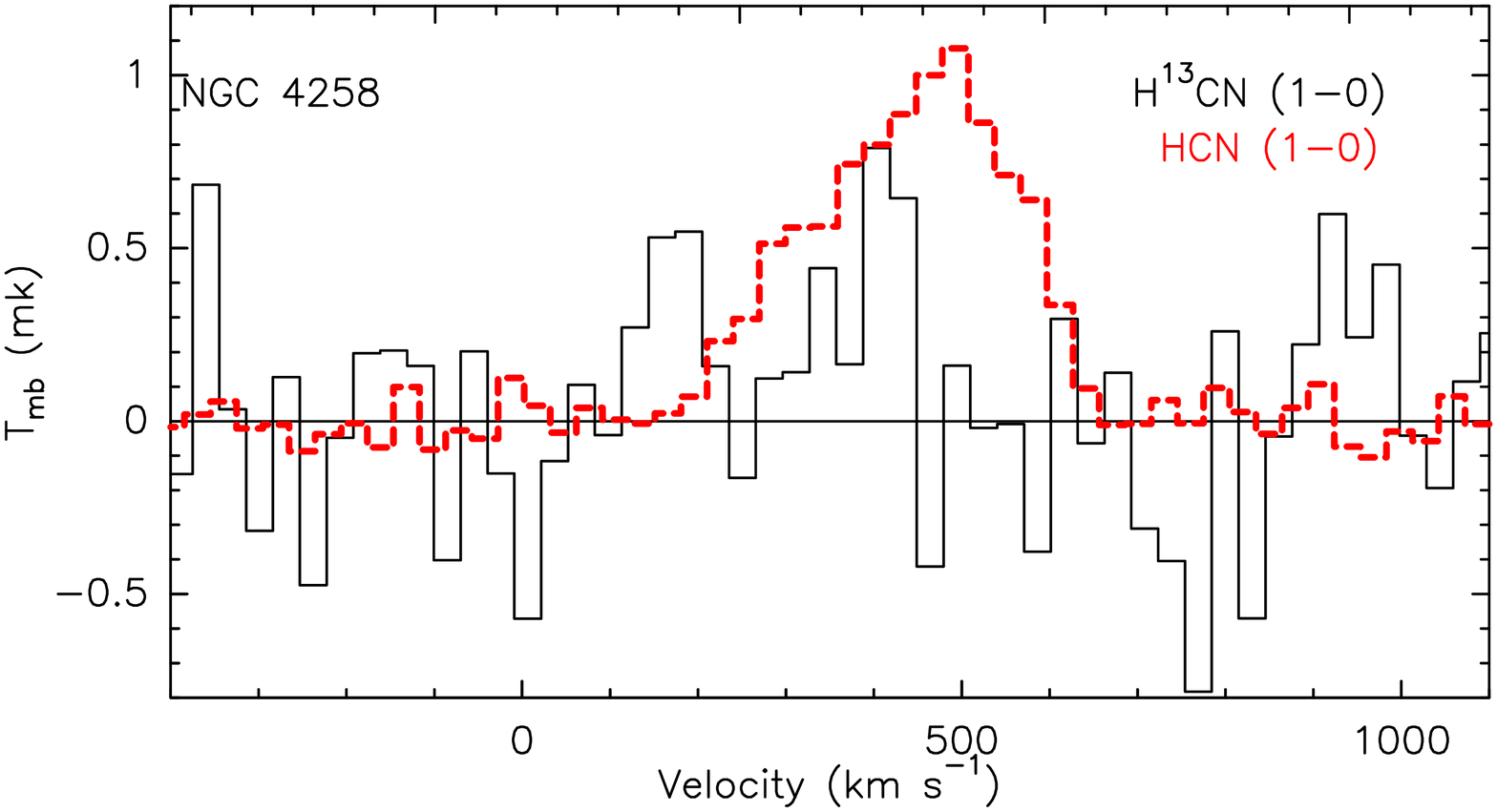}
\includegraphics[width=3in]{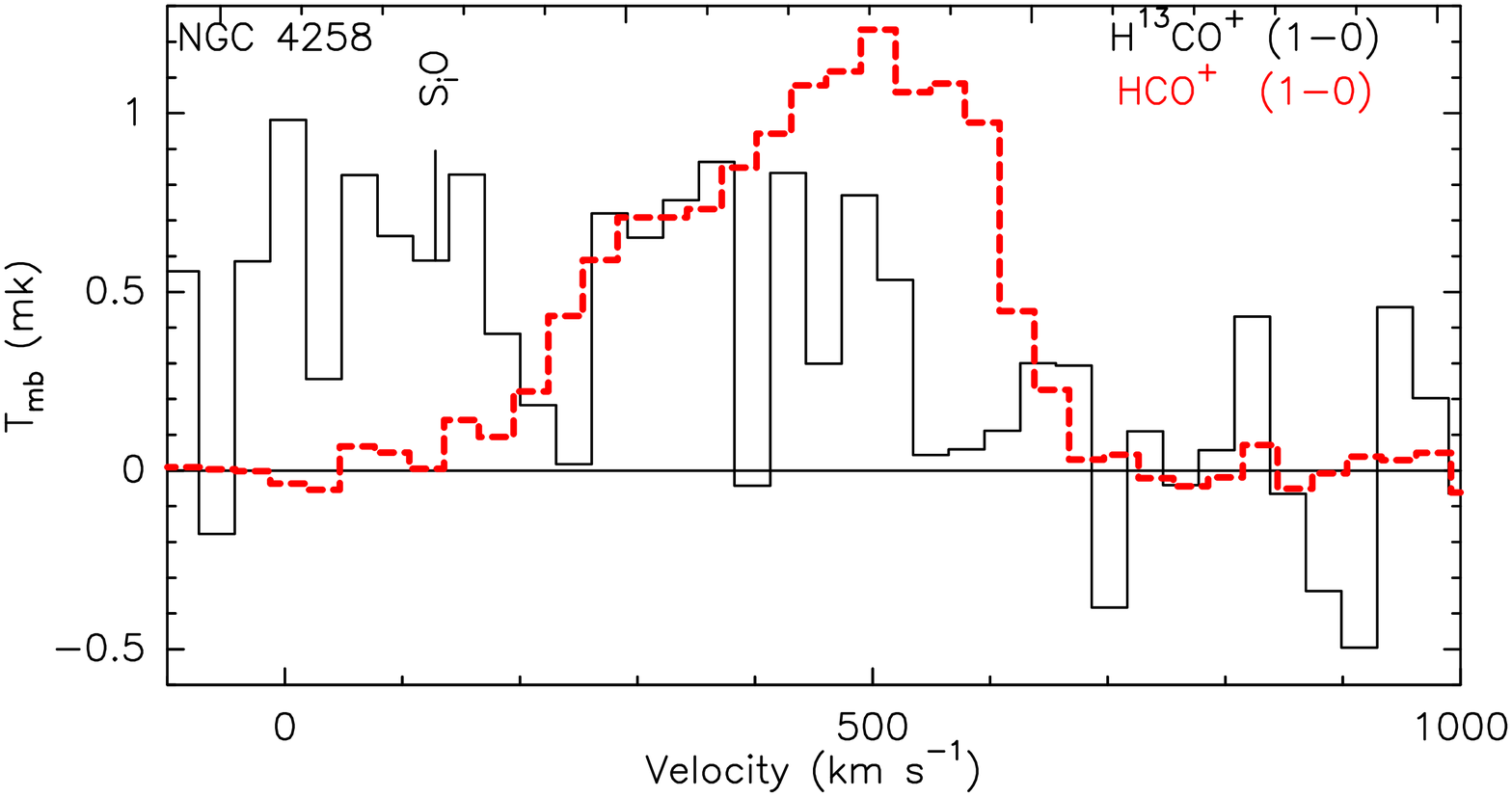}

\vspace*{-0.2 cm} \caption{Isotopologue lines of dense-gas tracers detected in NGC 4258. Left: H$^{13}$CN (1-0) (black line) overlaid with HCN (1-0) multiplied by 5 ( red line). 
Right: H$^{13}$CO$^+$ (1-0) (black line) overlaid with HCO$^+$ (1-0) multiplied by 7.2 ( red line). The SiO 2-1 line is also detected, which is at  left of  H$^{13}$CO$^+$ (1-0). 
	\label{fig:NGC4258_isotope}}
    \vskip-10pt
\end{figure*}

With better sensitivity than that of \cite{2011MNRAS.418.1753J},  we marginally detected H$^{13}$CO$^+$ and H$^{13}$CN for the first time above 3$\sigma$ level in NGC 4258. HCN 1-0 and HCO$^+$ 1-0 overlaid on their isotopic lines are presented in Figure \ref{fig:NGC4258_isotope}. With  assumption that  the isotopic ratios of molecules  are the same as the isotopic abundance of $^{12}$C/$^{13}$C  and this  ratio is  40, which was suggested for nearby galaxies \citep{1993A&A...268L..17H,2014A&A...565A...3H,2013A&A...549A..39A}, we can calculate the optical depths of HCO$^+$ 1-0 and  HCN 1-0, with  R = $\dfrac{1-e^{-\tau_{12}}}{1-e^{-\tau_{13}}}$, where R is the detected line intensity ratio and  ${\tau}$ is the optical depth. We obtained the optical depths for NGC 4258, and the upper limits for NGC 3079, respectively. The values of two sources has been showed in Table \ref{tabel 4}. The optical depths are similar to that in nuclear region of nearby type II Seyfert galaxy NGC 1068, with  the optical depths of HCN 1-0 and HCO$^+$ 1-0 as 4.6 and 3.0, respectively \citep{2014ApJ...796...57W}. The optically thick lines of HCN and HCO$^+$ can not well distinguish the abundance ratio of  HCO$^{+}$/HCN due to self absorption. Isotopic lines of dense gas tracers in galaxies can help us to derive the optical depths of dense gas tracers, which is important to accurate the relationship between the dense gas and the star formation \citep{2004ApJ...606..271G}, especially for individual galaxies. \cite{2008ApJ...677..262K} showed that IR  pumping is not significantly to effect the HCO$^+$-to-HCN intensity ratios. Line ratio of optically thin isotopologues (H$^{13}$CO$^+$/H$^{13}$CN),  which is more related to the relative abundances than that of HCO$^+$/HCN  with assumption of similar excitation temperatures,  is larger than unity in NGC 4258.


\begin{figure*}
	\centering
\includegraphics[width=3.2in]{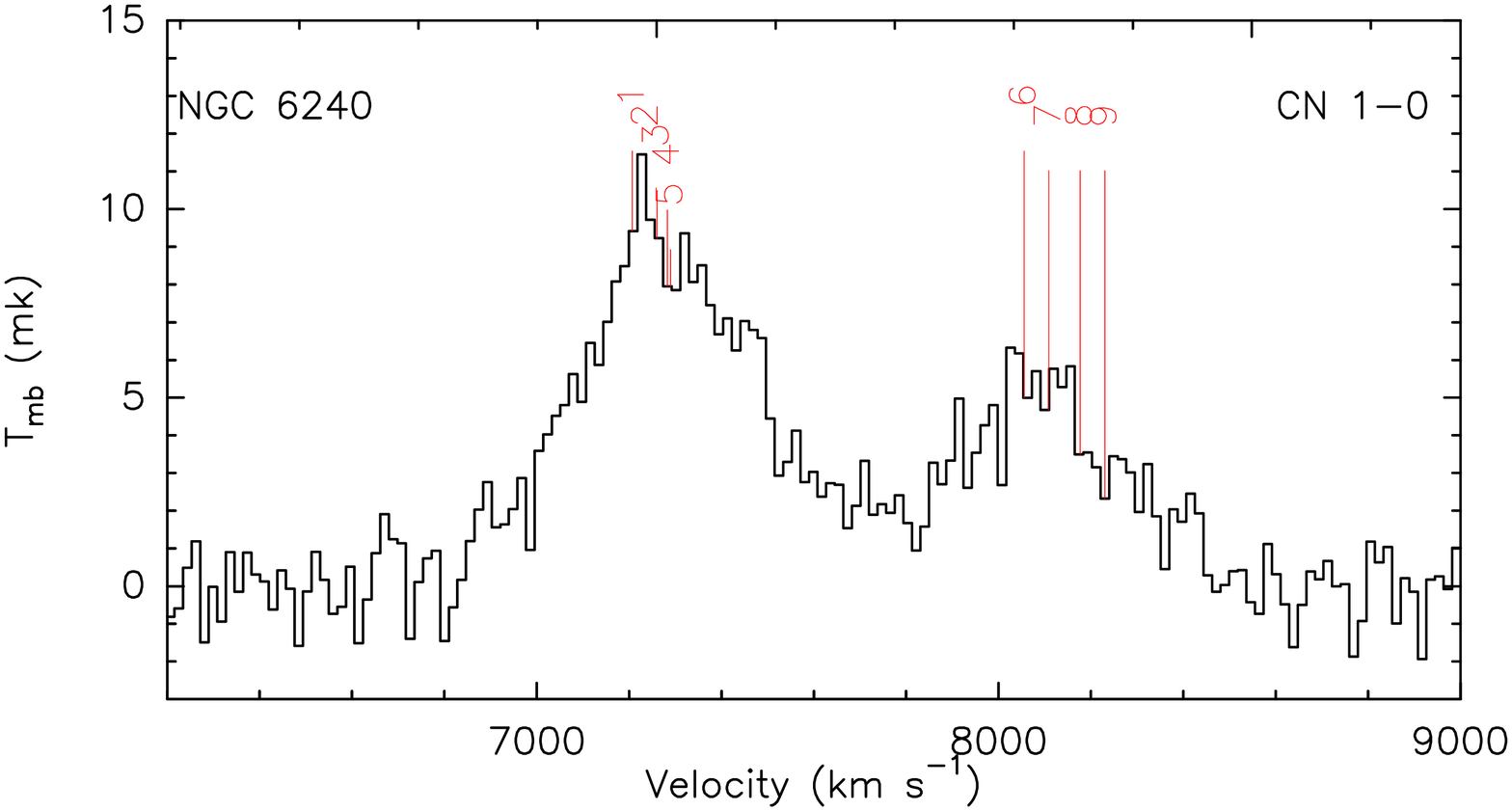}
\includegraphics[width=3.2in]{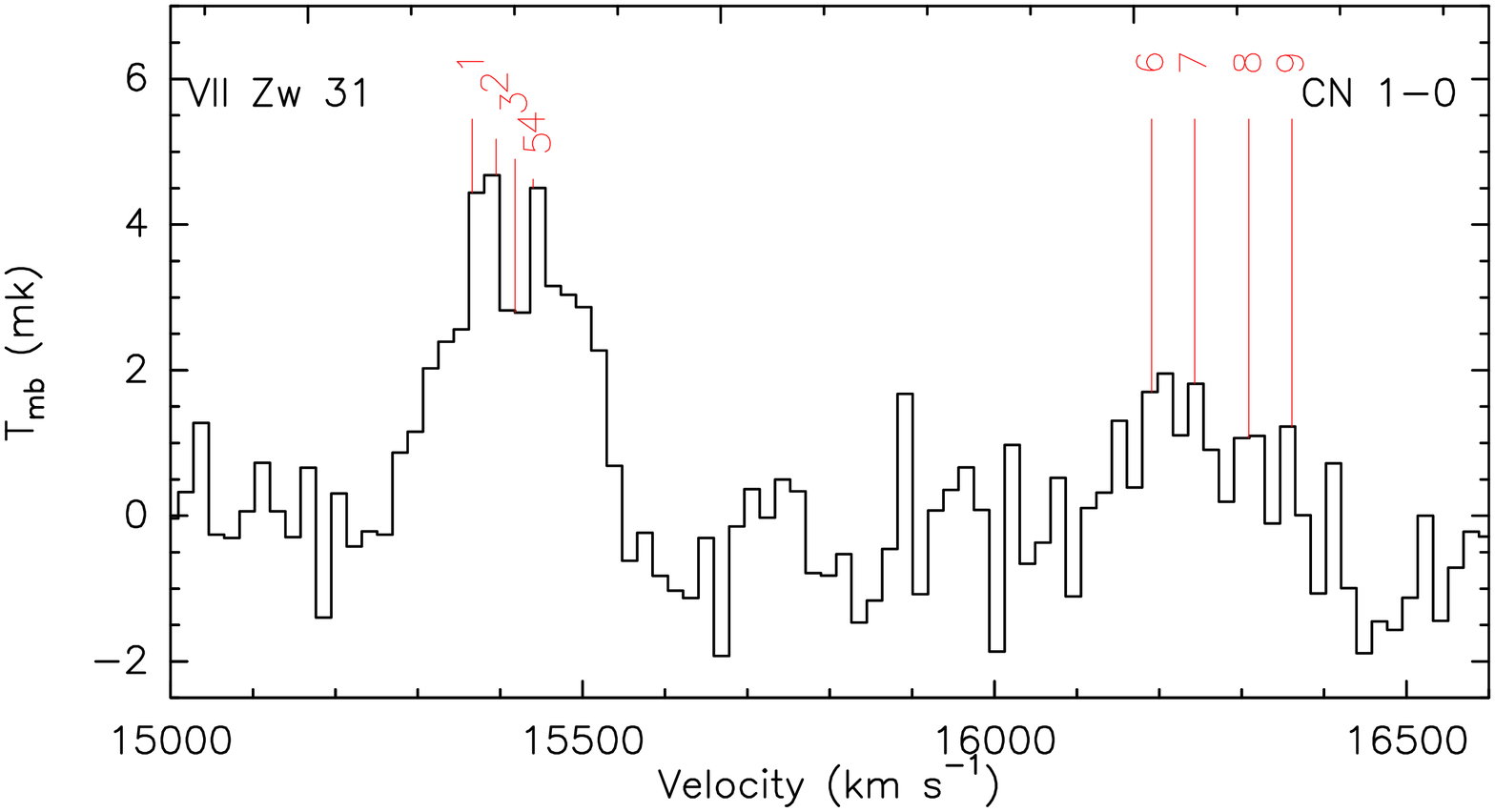}

\vspace*{-0.2 cm} \caption{CN (1-0) in NGC 6240 and VII ZW 31. The  rms  is  0.88 mK at the velocity resolution of 18.57 km\,s$^{-1}$ in NGC 6240 and 0.99 mK at the velocity resolution of 18.57 km\,s$^{-1}$. The same nine hyperfine lines  are marked from 1-9 in these two sources: 1, CN 1-0 J=3/2-1/2 F=1/2-3/2 at 113.520 GHz; 2, CN 1-0 J=3/2-1/2 F=3/2-3/2 at 113.509 GHz; 3, CN 1-0 J=3/2-1/2 F=1/2-1/2 at 113.4996 GHz; 4, CN 1-0 J=3/2-1/2 F=5/2-3/2 at 113.491 GHz; 5, CN 1-0 J=3/2-1/2 F=3/2-1/2 at 113.488 GHz;   6, CN 1-0 J=1/2-1/2 F=3/2-3/2 at 113.191 GHz; 7, CN 1-0 J=1/2-1/2 F=3/2-1/2 at 113.171GHz; 8, CN 1-0 J=1/2-1/2 F=1/2-3/2 at 113.144 GHz; 9, CN 1-0 J=1/2-1/2 F=1/2-1/2 at 113.123 GHz. The transition of J=3/2-1/2 F=1/2-3/2 in NGC 6240 and J=3/2-1/2 F=5/2-3/2 in VII Zw 31 is used as reference for the velocities, respectively.
	\label{fig:NGC6240_CN,VIIZW31_CN}}
    \vskip-10pt
\end{figure*}


\begin{table*}
\centering
 \begin{minipage}{160mm}
\caption{Derived optical depths for NGC 4258 and the upper limits for NGC 3079}
\label{tabel 4}
\begin{tabular}{l c c c c c c c c }
\hline
\hline
{Source} & {$\tau(HCN {1-0})$} & {$\tau(H^{13}CN {1-0})$}  & {$\tau(HCO^+ {1-0})$} & {$\tau(H^{13}CO^+ {1-0})$}  \\         
\hline
\hline
NGC 3079	&$<1.2 $ &	$<0.07 $ & $<1.0 $ & $<0.07 $\\	
NGC 4258	&$4.6 \pm 1.2$	 &	$ 0.12\pm0.03$   & $3.0 \pm 0.6$    & $0.08 \pm 0.02$ \\
		          
\hline	  
\end{tabular}
\end{minipage}
\end{table*}

\begin{table*}
\caption{Detected lines in NGC 3079}
\label{Table:NGC3079}
\centering
\setlength{\tabcolsep}{0.2in}
\begin{minipage}{160mm}
\begin{tabular}{l c c c c c c c c c c }
 \hline 
Molecule  &Frequency & $\int T_{mb}\ d\nu$	& $v_{LSR}$		&FWHM & Column densities\\
		& GHz		& (K km s$^{-1}$) & (km s$^{-1}$)	&(km s$^{-1}$)    &  ($\times10^{13}$cm$^{-2}$) \\
\hline

c-C$_3$H$_2$ \, $ (2_{1,2}-1_{0,1}  ) $    &  85.339     &  $  <0.11(3\sigma)^\mathrm{a}  $             &   . . .                      &  113.3                   &  $<2.7   $  \\          
                                                                    &                  &  $ 0.31 \pm 0.05^\mathrm{b} $               &   $1276.8\pm 10.3$  &   $113.3 \pm  20.3 $    & . . . \\  
H$^{13}$CN \, $ (1-0  ) $                            &  86.342  &  $  <0.25(3\sigma) $                &      . . .                        &       231.8                 &   $ <0.52$  \\
H$^{13}$CO$^+$ \, $ (1-0  ) $                    &  86.754  &  $  <0.22(3\sigma) $                &      . . .                         &     254.0                  &   $ <0.15  $\\

C$_2$H \, $(1-0)$                                      &   87.329  &  $ 3.54 \pm 0.17 $                   &   1117                           &   554.7                            &   $380\pm 18 $ \\
HNCO \, $ (4_{0,4}-3_{0,3}  ) $                  &  87.925   &  $  0.22\pm0.06^\mathrm{a} $ &  $876.3\pm13.7$      &   $103.9\pm37.3 $         &   $18 \pm 4.9 $ \\ 
                                                                   &                &  $  0.21\pm0.08^\mathrm{b} $ &  $1223.8\pm48.7$    &   $242.7 \pm81.4 $     &   . . .\\ 

HCN\,$ (1-0 ) $                                    & 88.631    &  $ 2.45\pm0.15^\mathrm{a} $  &  $987.2\pm6.7$        &   $205.7 \pm13.0 $      & $ 33\pm0.84 $\\ 
                                                                   &                &  $3.40\pm0.15^\mathrm{b} $   & $1277.3\pm 5.7$       &   $ 231.8 \pm11.5 $    & . . .\\       

HCO$^+$\,$ (1-0)$                             & 89.189    &  $2.64 \pm0.41^\mathrm{a} $   & $989.1  \pm16.9$      &   $223.7 \pm30.2$     &  $33\pm2.4 $ \\ 
                                                                   &                 &  $2.97\pm 0.41^\mathrm{b}$   & $1265.1\pm20.7$      &   $254.0 \pm32.8 $     & . . .  \\ 

HNC\,$ (1-0) $                                    &90.663      &  $0.52\pm 0.28^\mathrm{a} $   & $973.2 \pm 36.7$      &   $187.0 \pm97.3 $     &  $ 2.5\pm0.42$\\ 
                                                                   &                 &  $1.25 \pm 0.31^\mathrm{b} $   & $1257.1 \pm 20.2$    &  $243.8 \pm  68.9 $    &   . . .\\ 

HC$_3$N\,$ (10-9 ) $                                &90.979       &  $< 0.20(3\sigma)$                    &   . . .                           &     231.8                         &  $<1.3$  \\ 

C$^{18}$O\,$ (2-1  ) $                               &  219.560    &  $>1.32\pm 0.21$                       & $968.9\pm 12.1  $      & $139.6 \pm  21.8 $     & $ >138\pm22$ \\ 
$^{13}$CO\,$ (2-1  ) $                               & 220.399     &  $31.86 \pm 0.26  $                  & $1112.1  \pm 1.7   $    & $408.5 \pm3.4 $        &$3191\pm26 $\\ 
CH$_3$CN\,$ (12_{0}-11_{0}) $               &220.747      &  $0.95\pm 0.17$                       & $935.4  \pm 13.9  $     & $145.2 \pm  28.2 $    &$ 1.6\pm0.29$ \\ 

H$_2$CO\,$ (3_{1,2}-2_{1,1}  ) $             &225.698      &  $1.09\pm 0.19$                       & $860.9  \pm 14.7  $   & $172.7 \pm35.3 $      & $1.5\pm0.14 $  \\ 
                                                                 &                    & $0.98\pm 0.20$                        &$1208.0 \pm 18.3$     &$ 193.3 \pm 51.2  $    &. . .\\
CN\,$ (2-1) $                                            &226.660      &  $  2.06\pm 0.16^\mathrm{a}$   & $973.9\pm5.8  $          & $144.1 \pm11.7 $      & 
$103\pm  2$ \\ 
                                                                 &                   &  $0.66\pm 0.14^\mathrm{b}  $  & $1299.3\pm12.5  $       & $113.8 \pm  23.3 $    & . . . \\ 
CN\, $ (2-1) $                                           & 226.875     &  $  2.62\pm 0.20^\mathrm{a}  $ & $983.3\pm2.1$            & $193.2 \pm17.2 $      & . . . \\ 
                                                                 &                   &  $  3.47\pm 0.20^\mathrm{b}  $ & $1250.4  \pm1.4  $      & $180.2 \pm11.5 $      &  . . .\\

    \hline
\end{tabular}\\
Notes. The fluxes are in T$_{mb}$ instead of T$_A$. For C$_2$H, the hyperfine transition, the flux integrated from 514 km\,s$^{-1}$ to 1596 km\,s$^{-1}$. c-C$_3$H$_2$, HNCO, HCN, HCO$^+$, HNC and CN show a double pecked structure. $\mathrm{a}$ and $\mathrm{b}$ is the two Gaussian components. We use the red-shift component line width to estimate upper integrated intensity limit the blue-shift component of c-C$_3$H$_2$. We estimate upper integrated intensity limit of HC$_3$N, H$^{13}$CN with the red-shift component line width of HCN and H$^{13}$CO$^+$ with the red-shift component line width of HCO$^+$, respectively.  
 The column densities were calculated with the correction of an estimated source size.

\end{minipage}
\end{table*}

 \begin{table*}
\caption{Detected lines in NGC 4258}
\label{Table:NGC4258}
\centering
\setlength{\tabcolsep}{0.2in}
\begin{minipage}{160mm}
\begin{tabular}{l c c c c c c c c c c }
 \hline 
Molecule  &Frequency & $\int T_{mb}\ d\nu$	& $v_{LSR}$		&FWHM & Column densities\\
		& GHz		& (K km s$^{-1}$) & (km s$^{-1}$)	&(km s$^{-1}$)    &  ($\times10^{13}$cm$^{-2}$) \\
\hline

 c-C$_3$H$_2 $ \, $ (2_{1,2}-1_{0,1}  ) $     &   85.339  &  $0.11 \pm 0.03$   & $459.1 \pm 15.9$  &    $126.6\pm33.6$     & $2.4\pm0.65$ \\
H$^{13}$CN\,$ (1-0) $                                 &   86.342  &  $0.14 \pm0.04 $   & $391.4\pm 17.5$   &    $327.6 \pm93.2 $    & $ 0.79\pm0.23$\\ 
H$^{13}$CO$^+$\, $ (1-0) $                        &   86.754  &  $  0.17\pm0.05$   & $388.5\pm 40.2 $  &    $277.1\pm99.3 $   & $ 0.32\pm0.09 $ \\ 
SiO\,$ (2-1) $                                               &   86.847  &  $  0.12\pm0.04$   & $389.5 \pm 29.5$  &    $177.8 \pm50.0 $    & $ 0.65\pm0.22$ \\ 

C$_2$H\,$(1-0)$                                          &   87.329  & $0.96 \pm 0.07 $   & $459.9 \pm16.6$ &   $ 494.2 \pm 40.2 $   & $280\pm20$ \\
$^*$HCN\,$ (1-0) $                                      &   88.631  &  $1.38\pm0.04  $   & $ 451.0 \pm3.3$    &    $266.6 \pm7.3 $       & $21\pm0.61 $    \\  
$^*$HCO$^+$\,$ (1-0) $                                    &   89.189  &  $2.75 \pm0.04 $   & $459.1\pm2.4  $    &    $ 308.9 \pm5.3 $      & $ 89\pm1.3$\\ 

$^*$HNC\,$ (1-0) $                                      &   90.663  &  $0.40\pm0.04  $   & $482.9\pm10.4  $   &    $247.7 \pm  35.0 $   & $1.7\pm0.17$ \\ 

HC$_3$N\,$ (10-9) $                                  &   90.979  &  $0.11\pm0.03  $   & $561.9\pm 23.6 $   &    $232.1\pm94.8 $   & $ 1.6\pm0.44 $\\ 
C$^{18}$O\,$ (2-1) $                                  &  219.560   &  $<0.33(3\sigma)$    & . . .      &    162.5        & $<62 $ \\ 

$^{13}$CO\,$ (2-1) $                                  &  220.399 &  $3.79\pm 0.18$    & $495.0 \pm3.7  $    &    $162.5 \pm9.2 $       & $ 687\pm33 $ \\ 
CN\,$ (2-1) $                                              &226.660   &  $<0.17(3\sigma) $  &   . . .                        &  264.4                             &  $ <7.3$      \\
CN\, $ (2-1) $                                           & 226.875    &  $ <0.18(3\sigma)$  &   . . .                         &  264.4                              &    . . .  \\
$^*$CO\, $ (2-1) $                                      &  230.538 &  $ 80.90\pm 0.15$ & $485.6\pm0.2 $      &    $209.3 \pm0.5 $       & $ 14272\pm26$ \\ 

  \hline
\end{tabular}\\

The fluxes are in T$_{mb}$ instead of T$_A$. For C$_2$H, the hyperfine transition, the flux  integrated from -69 km\,s$^{-1}$ 1166 km\,s$^{-1}$. 
We use line width of HCN to estimate upper integrated intensity limit of CN. The asterisk $^*$ shows lines are optically thick and their column density would be underestimated.  The  column densities were calculated with the correction of an estimated source size. We estimate upper  limits of CN 2-1 and C$^{18}$O 2-1 integrated intensities with line width of HCN and $^{13}$CO, respectively.

\end{minipage}
\end{table*}

 \begin{table*}
\caption{Detected lines in NGC 6240}
\label{Table:NGC6240}
\centering
\setlength{\tabcolsep}{0.2in}
\begin{minipage}{160mm}
\begin{tabular}{l c c c c c c c c c c }
 \hline 
Molecule  &Frequency & $\int T_{mb}\ d\nu$	& $v_{LSR}$		&FWHM & Column densities\\
		& GHz		& (K km s$^{-1}$) & (km s$^{-1}$)	&(km s$^{-1}$)    &  ($\times10^{13}$cm$^{-2}$) \\
\hline

CN$^\mathrm{a}$ \,$ (1-0) $        &  113.191   &  $2.57\pm0.13$    &  $7229.2 \pm11.4$    &   $468.1\pm29.9 $   &  $3665\pm173$\\ 
CN$^\mathrm{b}$\,$ (1-0) $         &  113.491   &  $4.50\pm 0.13$   &  $7206.9 \pm6.5 $    &   $466.0 \pm16.1$   &  . . . \\ 
$^*$CO\,$ (1-0) $                         &  115.271   &  $60.41\pm0.42$  &  $7178.0 \pm1.3  $    &   $403.8 \pm3.4 $    &  $ 546166\pm3797 $ \\ 

C$_2$H \, $(3-2)$                        &   262.067  &  $4.28 \pm 0.25 $ &  $ 7144.0 \pm 10.6 $ &   $ 407.9\pm 30.6$  &  $ 130\pm7.6 $ \\

HC$_3$N \,$ (29-28) $                &   263.792  &  $0.28 \pm 0.06 $ &  $7174.7 \pm 5.9 $    &    $ 48.6 \pm11.6 $   &  $11\pm2.4 $  \\ 
HCN\,$ (3-2) $                      &  265.886   &  $9.22 \pm 0.31$  &  $7207.7 \pm 6.1$     &    $ 375.9 \pm14.5$  &  $ 7.8\pm0.26 $  \\ 
HCO$^+$\, $ (3-2) $             &267.558     &  $13.65\pm 0.66$ &  $ 7173.4 \pm9.4$     &    $ 397.4 \pm22.1$  &  $5.8\pm0.28 $ \\

  \hline
\end{tabular}\\

The fluxes are in T$_{mb}$ instead of T$_A$. For C$_2$H, the hyperfine transition, the flux integrated from 6799 km\,s$^{-1}$ to 7666 km\,s$^{-1}$. $^\mathrm{a}$ is one of the groups for CN at 113.191GHz. $^\mathrm{b}$ is the other group for CN at 113.491GHz. The asterisk $^*$ shows lines are optically thick and their column density would be underestimated. The  column densities were calculated with the correction of an estimated source size.

\end{minipage}
\end{table*}

 \begin{table*}
\caption{Detected lines in VII Zw 31}
\label{Table:VII Zw 31}
\centering
\setlength{\tabcolsep}{0.2in}
\begin{minipage}{160mm}
\begin{tabular}{l c c c c c c c c c c }
 \hline 
Molecule  &Frequency & $\int T_{mb}\ d$v	& v$_{LSR}$		&FWHM & Column densities\\
		& GHz		& (K km s$^{-1}$) & (km s$^{-1}$)	&(km s$^{-1}$)    &  ($\times10^{13}$cm$^{-2}$) \\
\hline

 CN$^\mathrm{a}$ \, $ (1-0) $        &  113.191   &   $0.30\pm0.10$   &  $15472.7 \pm 26.1$   & $158.0 \pm57.0 $   &   $ 129\pm43$ \\ 
CN$^\mathrm{b}$ \,$ (1-0) $         &  113.491   &   $0.75\pm0.06$   &  $15415.1 \pm7.5  $    & $169.2 \pm14.0 $   &  . . .    \\ 
$^*$CO\,$ (1-0) $                                 &  115.271   &    $17.53\pm0.06$ & $15440.3 \pm0.3  $     & $182.7 \pm0.7 $     &   $ 91210\pm312$ \\ 
$^*$CO\,$ (2-1) $                                 &  230.538   &    $41.81\pm0.15$ & $15442.6 \pm0.3  $     & $182.5 \pm0.7 $     &   . . . \\ 

C$_2$H\,$(3-2)$                           &  262.067   &    $0.47 \pm 0.10 $&  $15449.0 \pm18.5 $   & $175.3 \pm 39.5$   &    $ 884\pm188$\\
HC$_3$N (29-28)                         &263.792     &   $<0.3(3\sigma)$      & . . . &    178.2  &$<7.46$\\
HNCO\,$ (12_{1,11}-11_{0,10}) $ & 264.694    &    $0.47\pm 0.09  $& $15538.9 \pm 17.6  $  & $169.4 \pm33.7 $    &   $ 82\pm 19$ \\ 
HCN\,$ (3-2) $                              &265.886     &    $1.26\pm 0.14 $ & $ 15440.3 \pm 10.5  $ & $ 178.2 \pm20.1 $    &   $ 0.66\pm0.07 $  \\ 
HCO$^+$\, $ (3-2) $                     &267.558     &    $ 0.98\pm 0.15 $& $15443.5 \pm 15.2  $  & $198.6 \pm35.0 $     &$ 0.26\pm 0.04$  \\

  \hline
\end{tabular}\\

The fluxes are in T$_{mb}$ instead of T$_A$. For C$_2$H, the hyperfine transition, the flux integrated from 15250 km\,s$^{-1}$ to 15664 km\,s$^{-1}$.$^\mathrm{a}$ is one of the groups for CN at 113.191GHz. $^\mathrm{b}$ is the other group for CN at 113.491GHz. The asterisk $^*$ shows lines are optically thick and their column density would be underestimated. The   column densities were calculated with the correction of an estimated source size.

\end{minipage}
\end{table*}

\begin{table}
\begin {center} 
\setlength{\tabcolsep}{0.1in}
\caption{Column density ratios} 
\label{tabel 9}
\begin{tabular}{c c c c c c c c c c c}
\hline
\hline
{Sources} & $\dfrac{N_{H^{13}CN}}{N_{c-C_3H_2}}$&$\dfrac{N_{H^{13}CO^+}}{N_{c-C_3H_2}}$& $\dfrac{N_{HC_3N}}{N_{c-C_3H_2}}$\\         
  
\hline
\hline
NGC 3079  &	 $< 0.26   $           &	$<0.08$                &  $<0.65$             \\	
NGC 4258  &	 $0.33\pm 0.11$      &	$0.12\pm 0.053$      &  $0.67 \pm 0.30$  \\
NGC 6240  &	. . .	       &	. . .    &  . . .      \\
VII Zw 31	&	. . .	       &	. . .    &  . . .        \\
M 82         & $ 0.038 \pm 0.0049$  & $0.051 \pm 0.0032$  &$0.35 \pm 0.10$    	\\          
\hline	  
\end{tabular}
\end{center}
Notes. The column densities of M 82 were obtained from the literature: \cite{2015A&A...579A.101A}. The upper limit ratios were calculated in NGC 3079 by only the blue-shift component of c-C$_3$H$_2$. The column density of blue-shift component of c-C$_3$H$_2$ is 2.0*10$^{13}$ cm$^{-2}$.
\end{table}

\subsection{NGC 6240 v.s. VII Zw 31: AGN and extreme starburst hybrid v.s. extreme starburst}
\label{sec:SB}

NGC 6240 and VII Zw 31 are  about ten times more distant than NGC 3079 and NGC 4258. The IRAM beams covered the entire galaxy for both NGC 6240 and VII Zw 31.


\subsubsection{Molecules }

\label{sec:NGC6240_Molecular}
The detected lines in these two sources are similar, including CN 1-0, CO 1-0, C$_2$H 1-0, HCN 3-2, and HCO$^{+}$ 3-2, except that  HC$_3$N 29-28 line was only detected in NGC 6240 and HNCO 12-11 was  only detected in VII Zw 31. Although we did not detect new extragalactic molecules in these two sources, HNCO and HC$_3$N  29-28 was detected for the first time in VII Zw 31 and NGC 6240, respectively.  The detected species are listed in Tables \ref{Table:NGC6240} and \ref{Table:VII Zw 31}. Figure \ref{fig:colum_6240-zw31} show the comparison of molecular abundances among these two galaxies. Two groups of CN 1-0 lines, which include nine hyperfine transitions,  were detected both in NGC 6240 and VII Zw 31. Due to the line  broadening of galaxies, the nine CN 1-0 lines were in two groups, five of which were around 113.491 GHz while four of which were around 113.191 GHz. The detected CN$_{113.491 GHz}$ / CN$_{113.191 GHz}$ ratios in  NGC 6240 and VII Zw 31 were about 2,  which were similar to the optical depth ratio of the two groups and  indicated that the lines were optically thin. C$_2$H 3-2 was detected in these two sources. However,  due to the line broadening, the 4 hyperfines can  not be resolved (See Figure \ref{fig:NGC6240_C2H,VIIZW31_C2H}).

Some transitions of HC$_3$N in galaxies  had been detected  \citep{2011A&A...525A..89A,2011A&A...528A..30C,2011A&A...527A.150L,Jiang17}, with low detection rate. So far, HC$_3$N 32-31 in NGC 4418 was the highest transition detected in external galaxies \citep{2015A&A...582A..91C}. With the detection of HC$_3$N 29-28, which requires the excitation conditions with high density (around $10^6$ cm$^{-3}$) and high temperature ($E_u\sim$190 K) \citep{2015A&A...579A.101A}, at least part of  molecular gas in  NGC 6240 should have  similar physical properties  to that in NGC 4418. HNCO, as a good tracer of the evolution of nuclear SB in galaxies and a shock tracer \citep{2015A&A...579A.101A,2008ApJ...678..245M}, was detected in VII Zw 31, which implied that the ISM of VII Zw 31 might  be  dominated by shocks affecting the molecular clouds fueling the SB.

\subsubsection{Line ratios}

The line ratio of  HCO$^+$/HCN 3-2 was 1.5 in NGC 6240,  which was about twice of  the value in \cite{2008ApJ...677..262K}. HCO$^+$ and HCN data from the observation at different time in \cite{2008ApJ...677..262K}, with HCO$^+$ data from \cite{2006ApJ...640L.135G}. The velocity integrated  intensity of HCN 3-2 in our results  was almost the same as that in \cite{2008ApJ...677..262K}, while it was more than twice flux in our results than that of  \cite{2008ApJ...677..262K} for HCO$^+$. The pointing error can underestimate flux of HCO$^+$,  resulting in underestimating line ratio of HCO$^+$/HCN. The ratio was 0.78 for VII Zw 31, as a typical value for ULIRGs \citep{2011A&A...528A..30C}. While the higher HCO$^+$/HCN 3-2 ratio in NGC 6240 is close to SB-dominated source, as seen in starburst-dominant galaxy nuclei. With two prominent AGN in NGC 6240, we suggest that an enhanced HCN abundance in an AGN could be the reason for the higher HCN flux, resulting in higher HCO$^+$/HCN 3-2 in NGC 6240 than VII Zw 31, in which no obvious AGN signature exists.


 

\subsection{Comparison with observations in literature}

The derived ratios of HCO$^+$/HCN agree well with previous studies for NGC4258 \citep{2011MNRAS.418.1753J} and NGC 3079 \citep{2008ApJ...677..262K,2011A&A...528A..30C}, while it is about  twice for NGC 6240 than that in \cite{2008ApJ...677..262K}.  HCO$^+$/HCN  3-2 ratio is a new measurement for VII Zw 31. More HCO$^+$/HCN ratios for other sources from the literature are listed in Table \ref{tabel 3}, which  show that starbursts have, on average, higher HCO$^+$/HCN ratios than AGN. The HCO$^+$/HCN ratio can be used to observationally distinguish AGN-important and starburst-dominant galaxy nuclei.  For the composite sources (NGC 6240 and NGC 3079), we found that this ratio is moderately low comparing to starburst galaxies. We suggest that the line intensity may be  contaminated from coexisting AGN region have a lower HCO$^+$/HCN ratio. This indicates that the HCO$^+$/HCN may not be a perfect diagnostic tool between AGN and SB environments, due to combination of both processes. 


\subsection{HC$_3$N/HCN}

Table \ref{tabel 3} show upper limit of HC$_3$N/HCN ratio from literature. The starburst galaxies are, on average, lower HC$_3$N/HCN ratios comparing with  AGNs. It is consistent with \cite{2011A&A...527A.150L}, which concluded that starburst galaxies seem to be poor in HC$_3$N, because HC$_3$N is easily dissociated by UV radiation. As mentioned in Sec. \ref{sec:NGC6240_Molecular}, HC$_3$N 29-28 was detected in NGC 6240 with the ratio of HC$_3$N 29-28/HCN 3-2 of 0.03, which gave that HC$_3$N abundance of NGC 6240 was lower than that of other two ULIRGs: Arp 220 and Mrk 231 \citep{2015A&A...579A.101A}. With the extreme starburst activity, we suggest that NGC 6240 also belongs to the HC$_3$N-poor galaxy, which is consistent with the result of \cite{Jiang17} with non-detection of HC$_3$N emission. The upper limit ratio of HC$_3$N/HCN in NGC 3079 is 0.03, which is consistent with \cite{2011A&A...527A.150L} as an HC$_3$N-poor galaxy. The non-detection HC$_3$N line in NGC 3079 and VII Z w 31 and $\dfrac{I(HC_3N)}{I(HCN)}$ $<$ 0.15 in NGC 4258 and NGC 6240, which implies that all four galaxies are HC$_3$N-poor galaxies.

 \setlength{\parskip}{1.2\baselineskip}
\subsection{Abundances of several molecules}
\label{sec:abundances}

In our sample, NGC 3079 and NGC 6240 host both AGN and  powerful SB,  while NGC 4258 is a pure AGN  and  VII Zw 31 is an ULIRG mainly powered by SB.  C$_2$H, as a good PDR tracer,  has similar abundances in NGC 3079, NGC 4258 and NGC 6240, which is an order of magnitude lower than that in VII Zw 31, indicating that strong  UV fields in VII Zw 31 might be required for the enhancement of  C$_2$H. $c$-$C_3H_2$  line is tentatively detected in NGC 3079 and NGC 4258. However, CN, whose abundance should also be enhanced in an X-ray chemistry \citep{2007A&A...461..793M}, was not detected in NGC 4258. High resolution observations are needed to better understand CN and $c$-$C_3H_2$ emission in galaxies with AGN and/or starburst.

Column density ratios of  molecules with less affect of opacity effects, were listed in Table  \ref{tabel 9}. $\dfrac{N_{H^{13}CN}}{N_{c-C_3H_2}}$, $\dfrac{N_{H^{13}CO^+}}{N_{c-C_3H_2}}$ and $\dfrac{N_{HC_3N}}{N_{c-C_3H_2}}$ in NGC 3079,  as non-detections, were lower  than that in NGC 4258, which means that the abundances of H$^{13}$CN,  H$^{13}$CO$^{+}$, and HC$_3$N molecules are relatively lower in NGC 3079 than that in NGC 4258, if a fixed abundance of ${c-C_3H_2}$ is assumed in both galaxies. The ratios of $\dfrac{N_{H^{13}CO^+}}{N_{c-C_3H_2}}$, $\dfrac{N_{H^{13}CN}}{N_{c-C_3H_2}}$ and $\dfrac{N_{HC_3N}}{N_{c-C_3H_2}}$ in NGC 4258 are  higher than that  in M 82 \citep{2015A&A...579A.101A}, which may be caused by the enhancement of  $c$-$C_3H_2$ in M82 \citep{2011A&A...535A..84A}.

\section{Summary and Conclusions}

We presented  3mm and 1mm band observations  toward four nearby galaxies, including NGC 3079, NGC 4258, NGC 6240 and VII Zw 31,  with the IRAM 30m telescope. The main results of this study are as follows.

1. We detected 10 molecular species in NGC 3079, 11 molecular species in NGC 4258, 6 molecular species in NGC 6240, and 6 molecular species in VII Zw 31, which expanded the molecular line data set for these galaxies. 
HC$_3$N 29-28 which requires high excitation condition, was detected in NGC 6240, and HNCO 12-11 was detected for the first time in VII Zw 31. In addition, we tentatively detected the isotopic lines  $H^{13}CN$ 1-0 and $H^{13}CO^+$ 1-0 about 3$\sigma$ level in NGC 4258. Optical depths of  dense gas tracers were estimated as 4.1 for HCN 1-0 and 2.7 for HCO$^{+}$ 1-0 in NGC 4258 with the detected  $H^{13}CN$ 1-0 and $H^{13}CO^+$ 1-0 lines.

2. We performed the detailed comparison of  two nearby type II AGN: NGC 3079 and NGC 4258; and two gas-rich (U)LIRGs: VII Zw 31 and NGC 6240, with detected lines. Two groups of CN 1-0  in VII Zw 31 and  NGC 6240,  were detected and estimated as optically thin lines based on the line ratio. The high line ratio of HCO$^{+}$/HCN 1-0 in NGC 4258 is inconsistent with typical value of pure AGNs and composite sources, while the line ratio of $H^{13}CO^+/H^{13}CN$ was close to unity, which is more relate to the relative abundances. The intensity ratio of HCO$^+$/HCN lines might not  be an ideal tool to identify AGN or starburst in galaxies. The ratio of $^{13}$CO/C$^{18}$O 2-1 shows obvious difference in NGC 3079 and NGC 4258.


3. Under the LTE approximation, we derived column densities of various molecular species. With the relative abundances of C$_2$H, HNCO, HCN, HCO$^+$ and HNC normalized by CN, we found the environment of starburst in NGC 3079 might be similar to that in NGC 253. c-C$_3$H$_2$ was detected, while CN were not detected, in NGC 4258. CN molecule shows abundance variations among different power sources of galaxies nuclei.

4. We find these four source are all HC$_3$N-poor galaxies. $\dfrac{I(HC_3N)}{I(HCN)} $ $<$ 0.15 in NGC 4258 and NGC 6240 and the non-detection HC$_3$N line in NGC 3079 and VII Z w 31.

\section*{Acknowledgements}

We thank useful comments and suggestions from the anonymous referee. 
This work is supported by the National Key R$\&$D Program of China (No. 2017YFA0402704),  the
Natural Science Foundation of China under grants of 11590783, the Youth Foundation of Hebei Province of China (No. A2011205067) and National Youth Fund (No. 11303008) and Astronomical Union Fund (No. U1831126). We thank the staffs at the IRAM 30m telescope for their kind help and support during our observations.

{}

\appendix

\section{Molecular species in these sources}

\begin{figure*}
	\centering 
\includegraphics[width=3.2in]{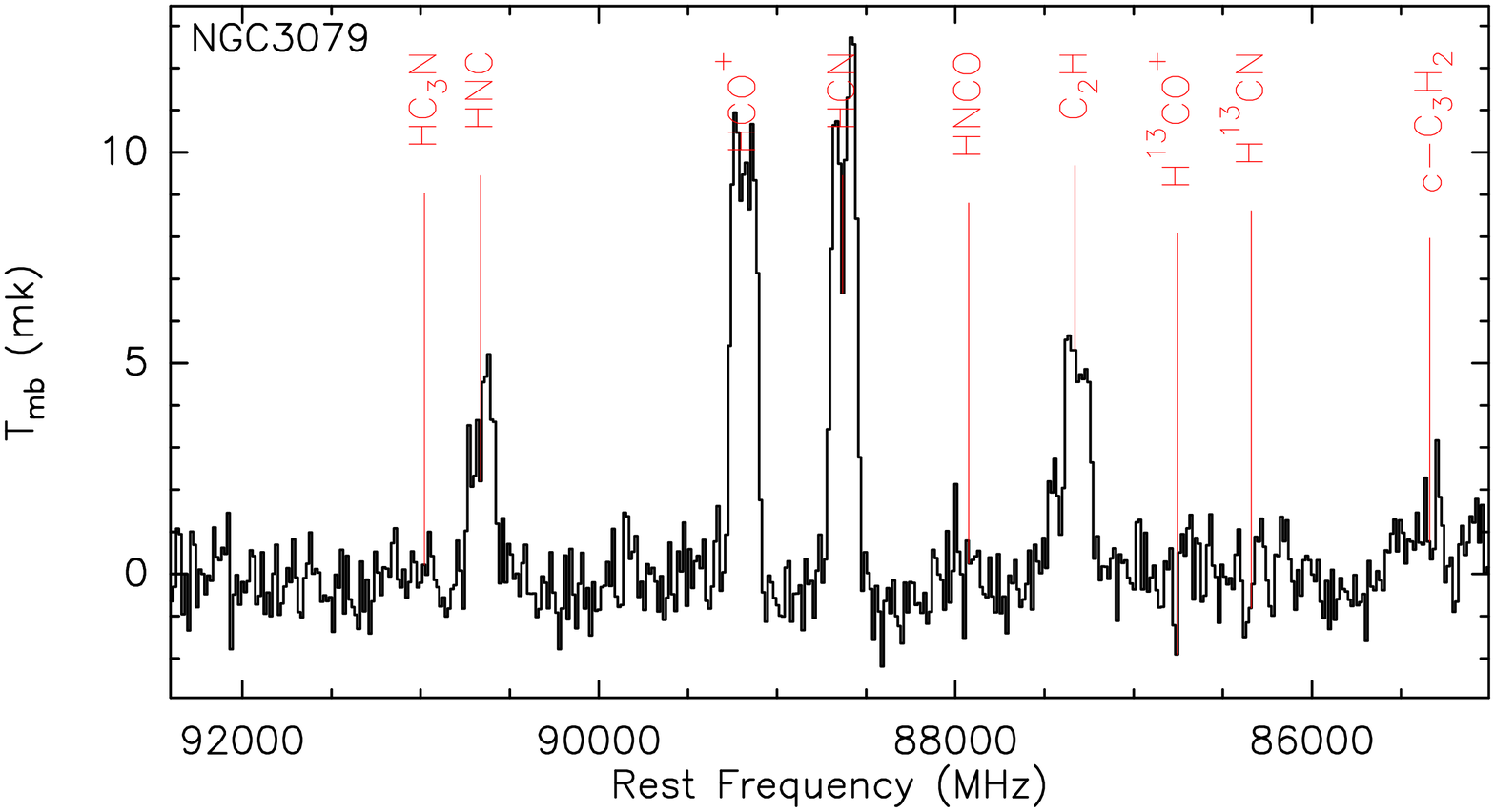}
\includegraphics[width=3.2in]{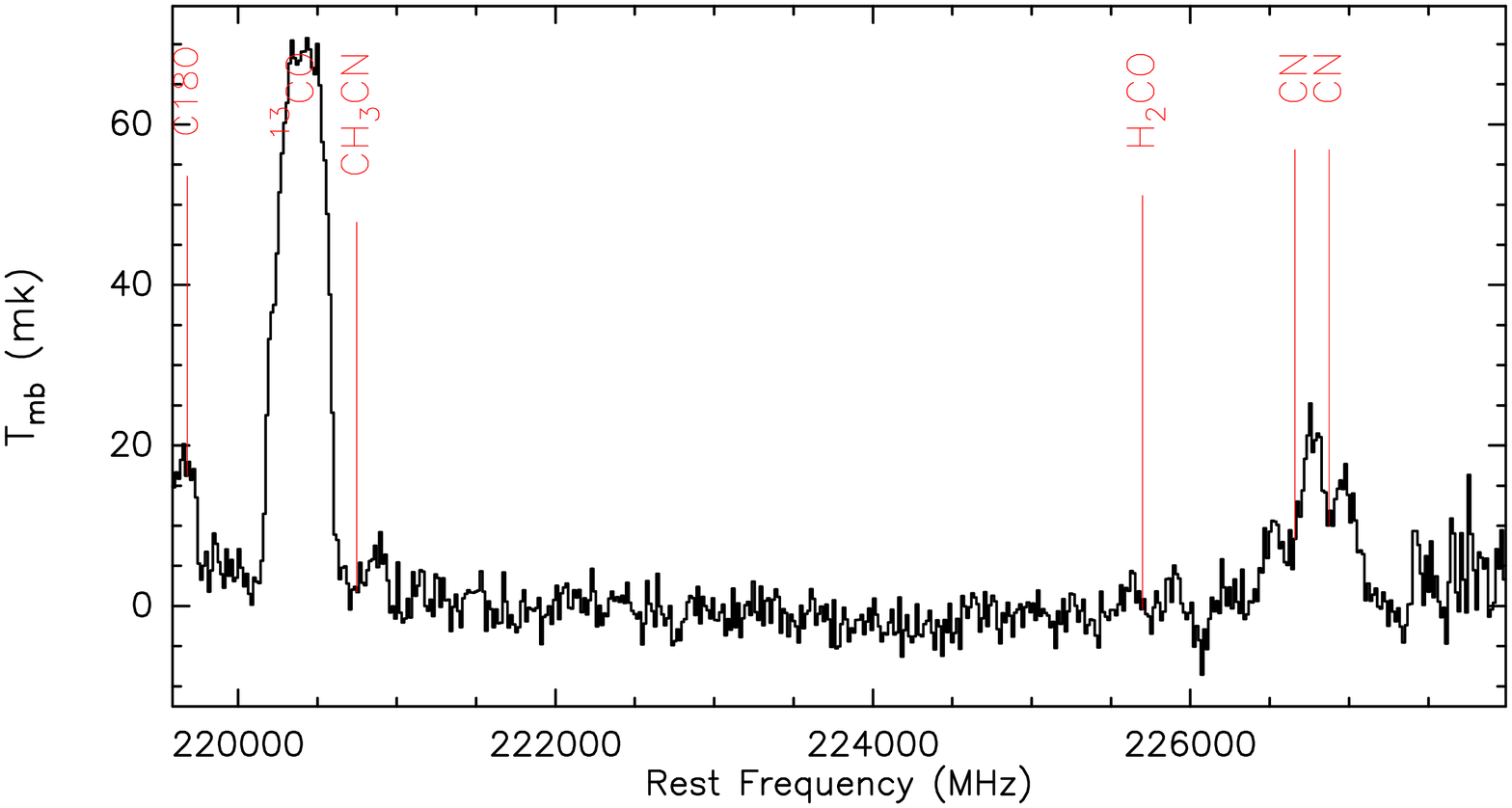}
\vspace*{-0.2 cm} \caption{Molecular species of NGC 3079, Left:band ranges from 85 GHz-92 GHz, with RMS of 1.13 mK at the velocity resolution of 54.67 km\,s$^{-1}$. 
Right: band ranges from 220 GHz-227 GHz, with RMS of  3.29 mK at the  velocity resolution  of  21.42 km\,s$^{-1}$. 
	\label{fig:NGC3079}}
    \vskip-0.5pt
\end{figure*}


\begin{figure*}
	\centering 
\includegraphics[width=3.2in]{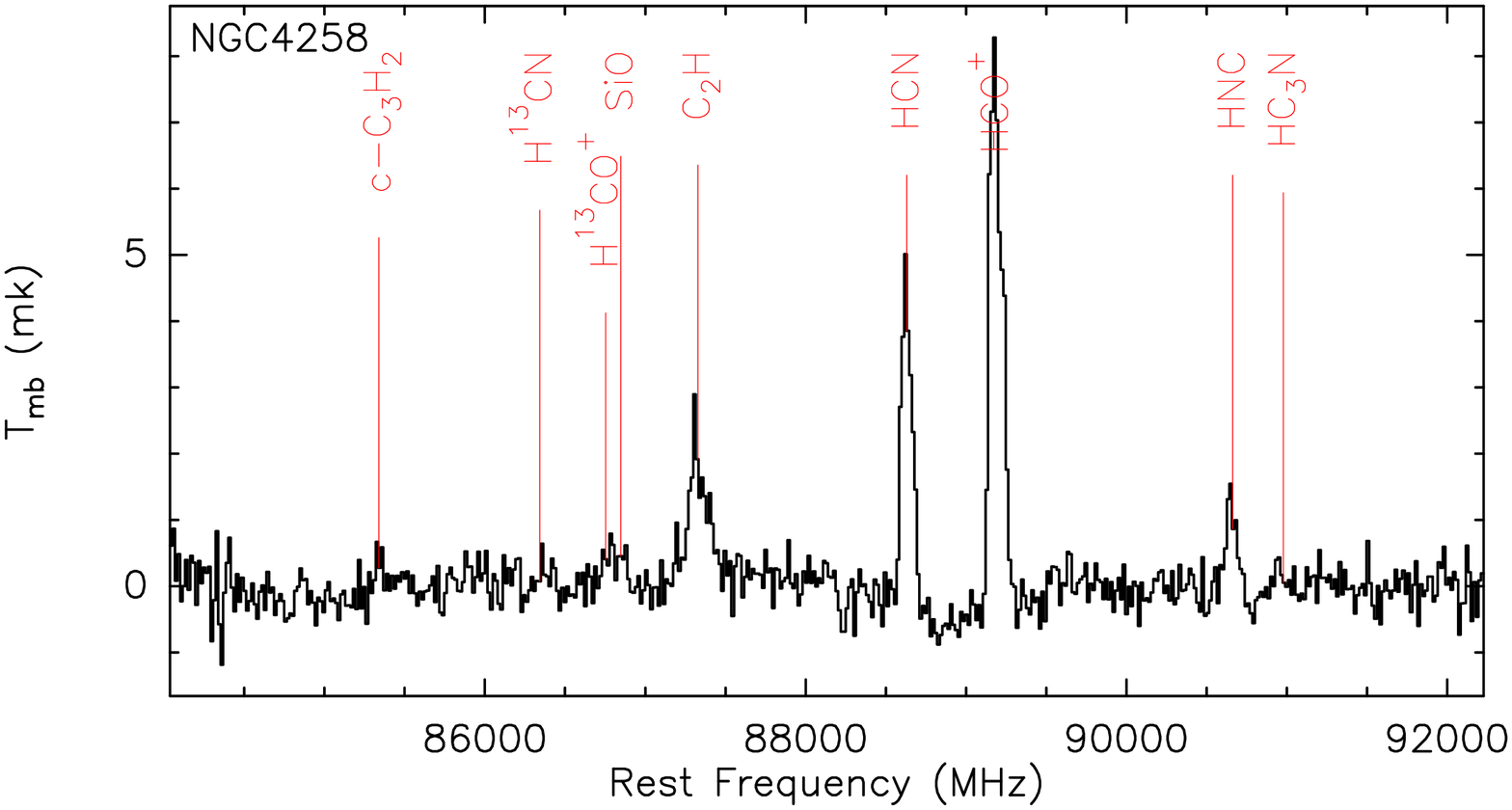}
\includegraphics[width=3.2in]{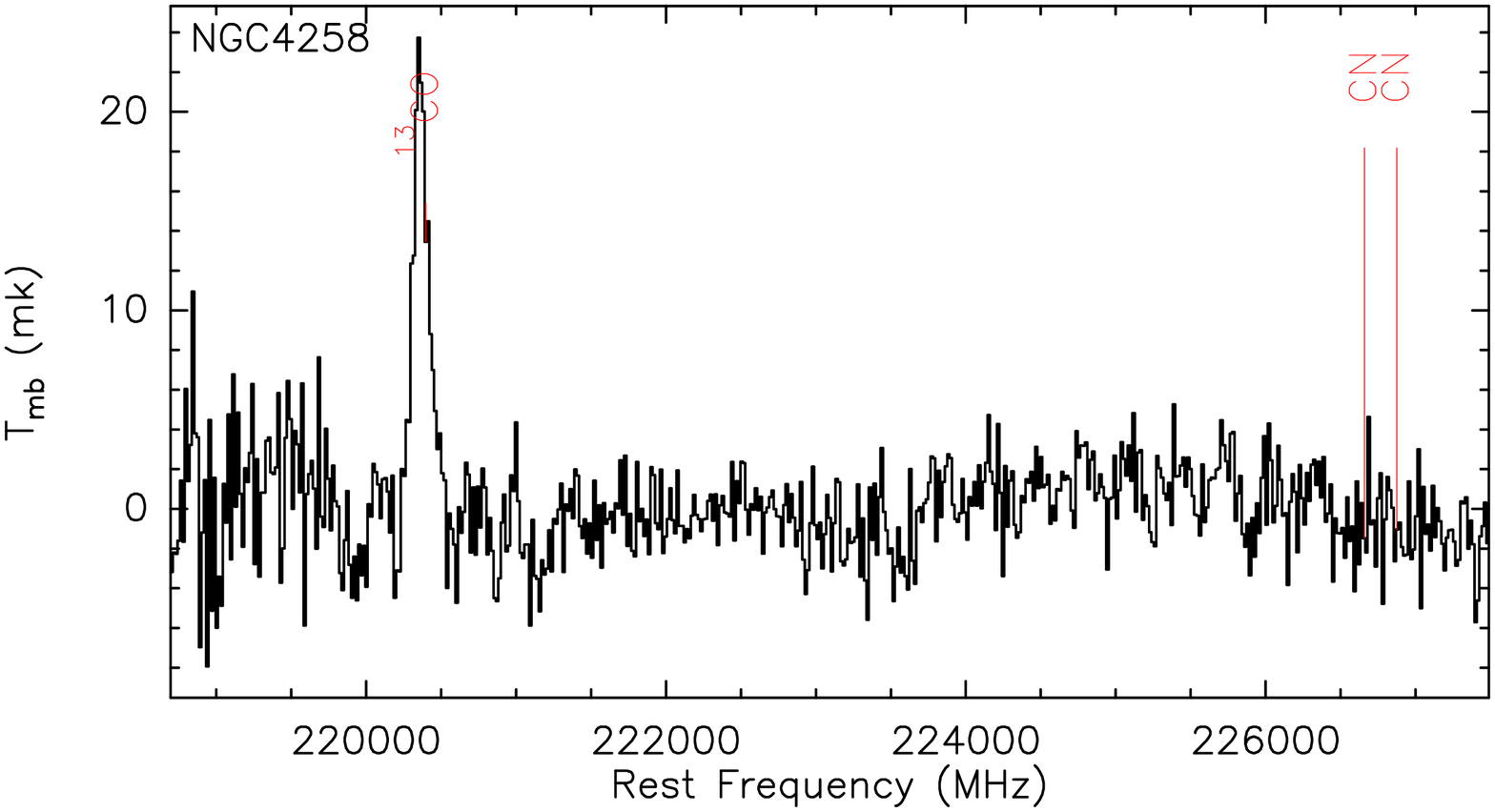}
\includegraphics[width=3.2in]{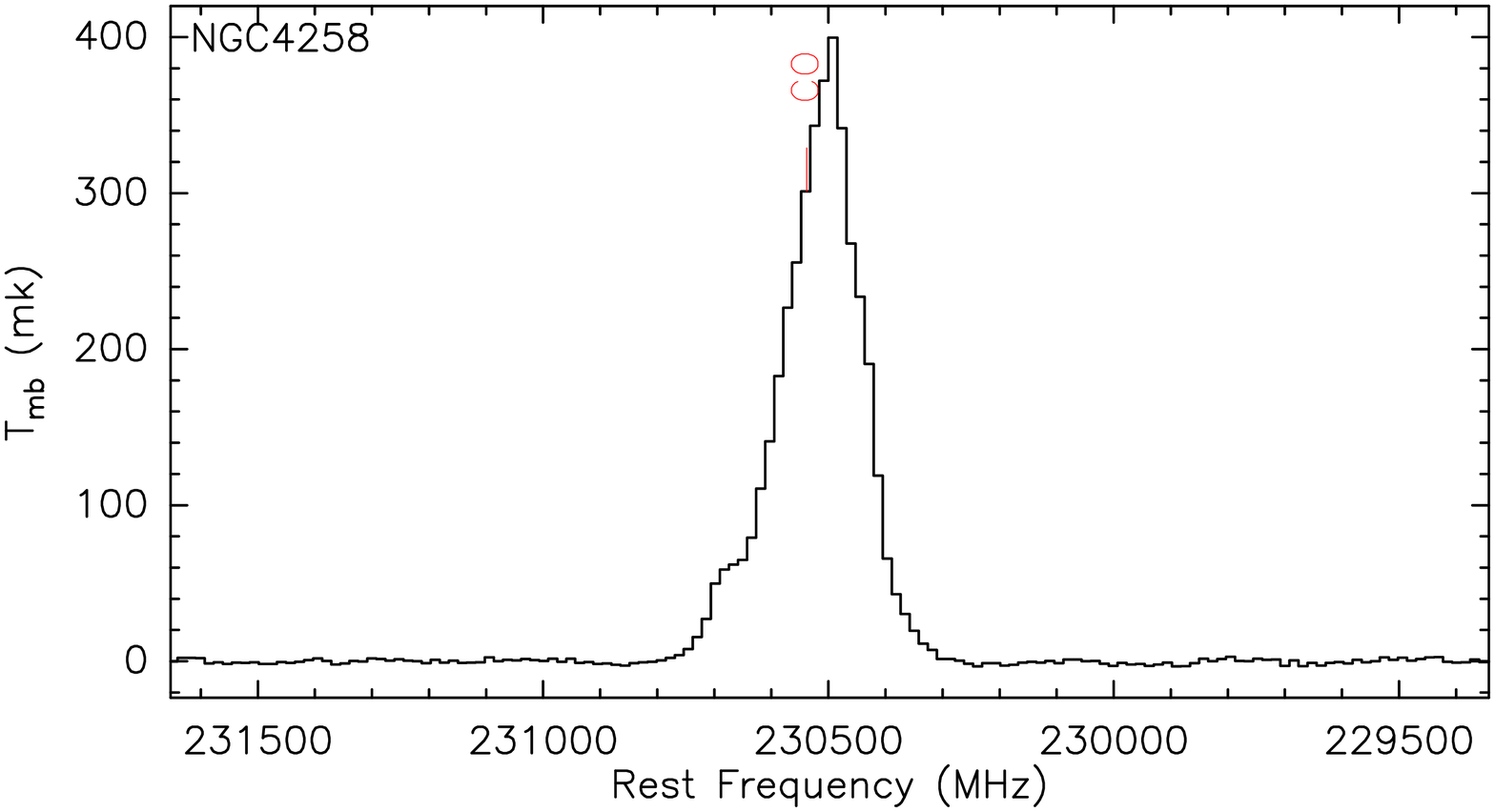}
\vspace*{-0.2 cm} \caption{Molecular species of  NGC4258. Top left:  band ranges from 85 GHz-91 GHz, with the  RMS of  0.33 mK  at the  velocity resolution of  -54.67 km$\cdot$s$^{-1}$.  
Top right: band ranges from 219 GHz-227 GHz, with the  RMS of 2.33 mK at the  velocity resolution of  -21.36 km$\cdot$s$^{-1}$. 
 Bottom: band ranges from 229 GHz-236 GHz, with RMS  of 1.25 mK  at the  velocity resolution of  20.57 km\,s$^{-1}$.  
	\label{fig:NGC4258}}
    \vskip-0.5pt
\end{figure*}

\begin{figure*}
	\centering 
\includegraphics[width=3.2in]{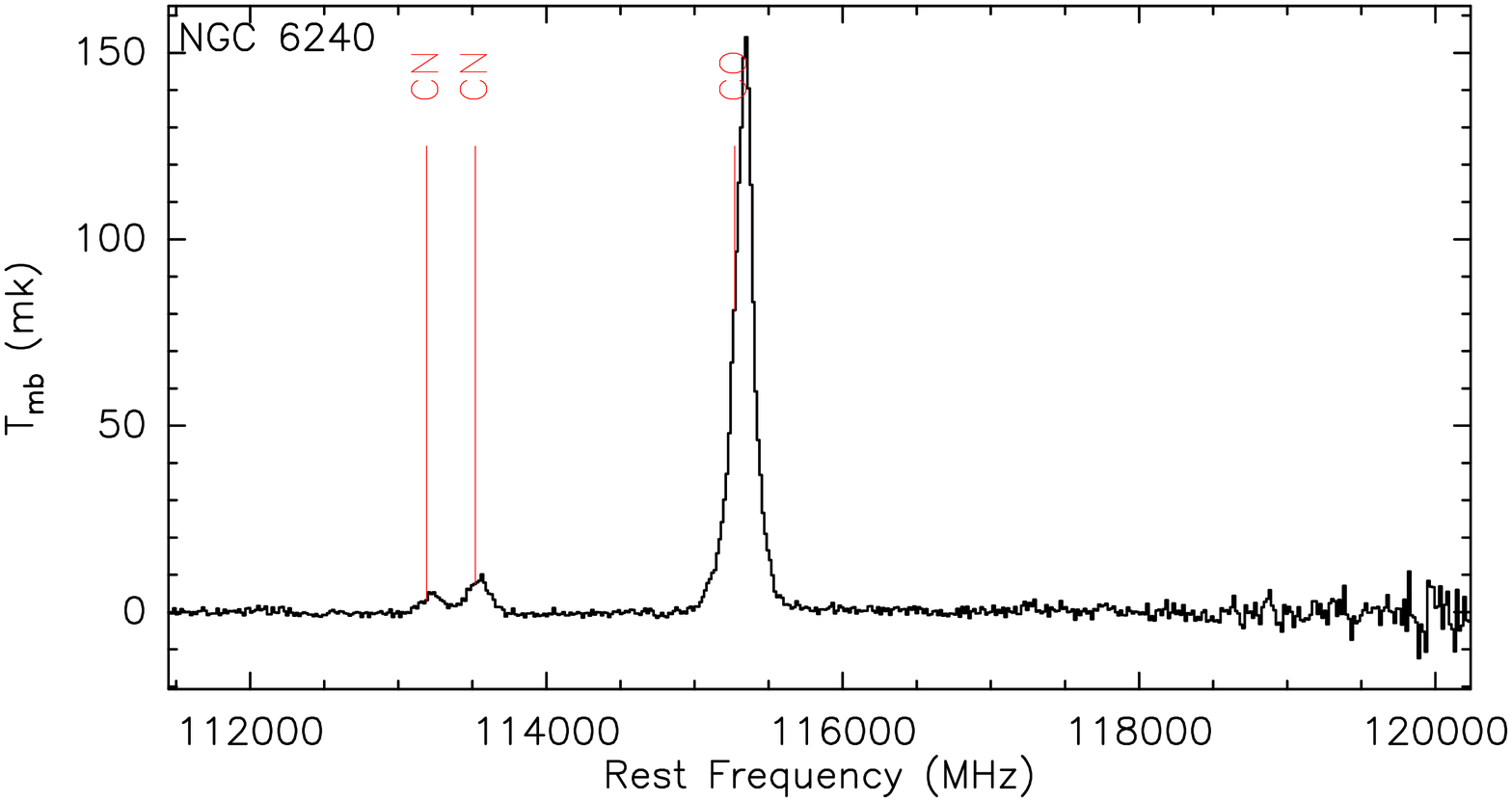}
\includegraphics[width=3.2in]{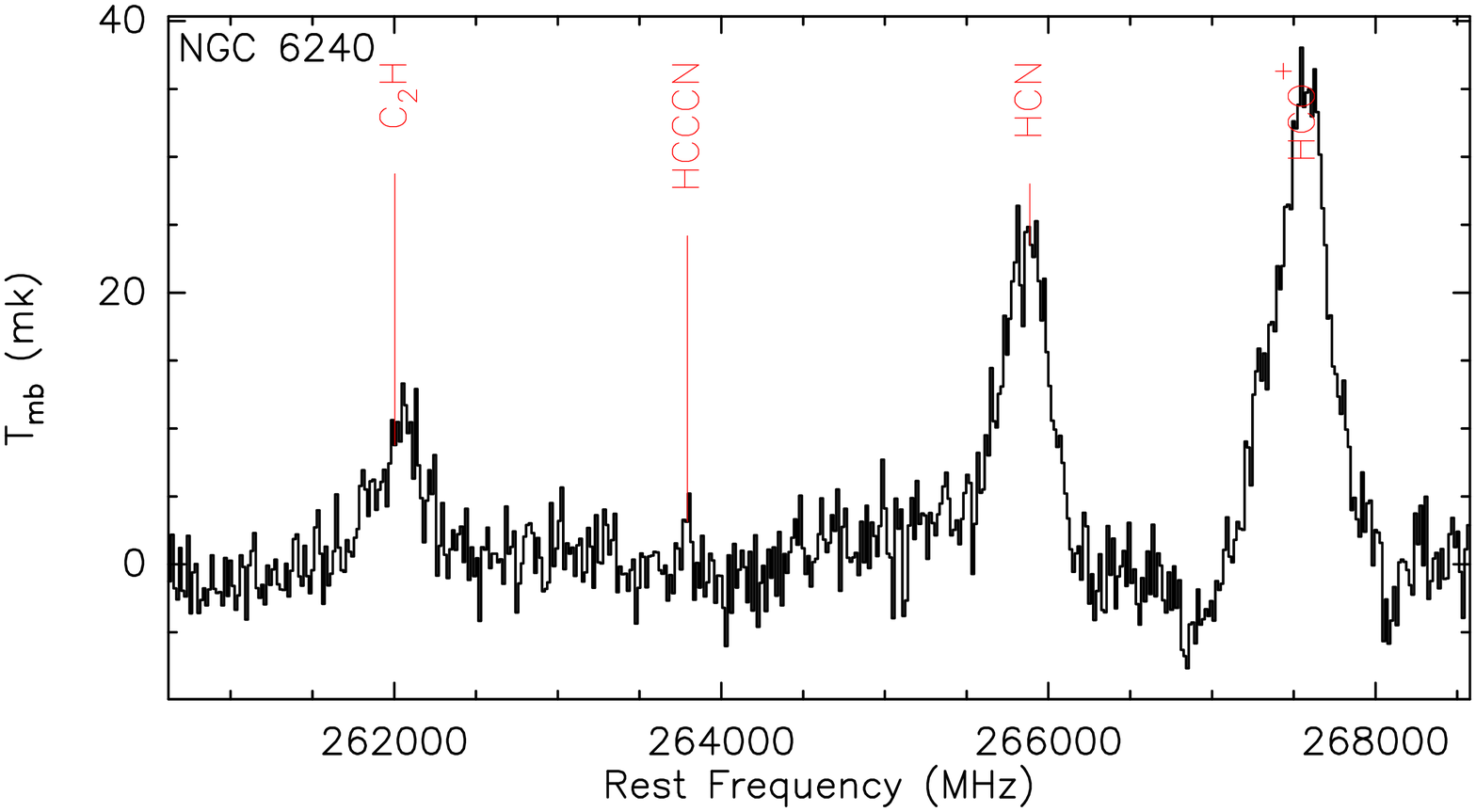}

\vspace*{-0.2 cm} \caption{Molecular species of the NGC 6240. Left: band ranges from 109 GHz-117 GHz,  with RMS of 2.03 mK at the  velocity resolution of 41.53 km\,s$^{-1}$. 
Right: band ranges from 255 GHz-262 GHz, with RMS of 2.40 mK at  the velocity resolution of 18.25 km\,s$^{-1}$. 
	\label{fig:NGC6240}}
    \vskip-0.5pt
\end{figure*}

\begin{figure*}
	\centering
\includegraphics[width=3.2in]{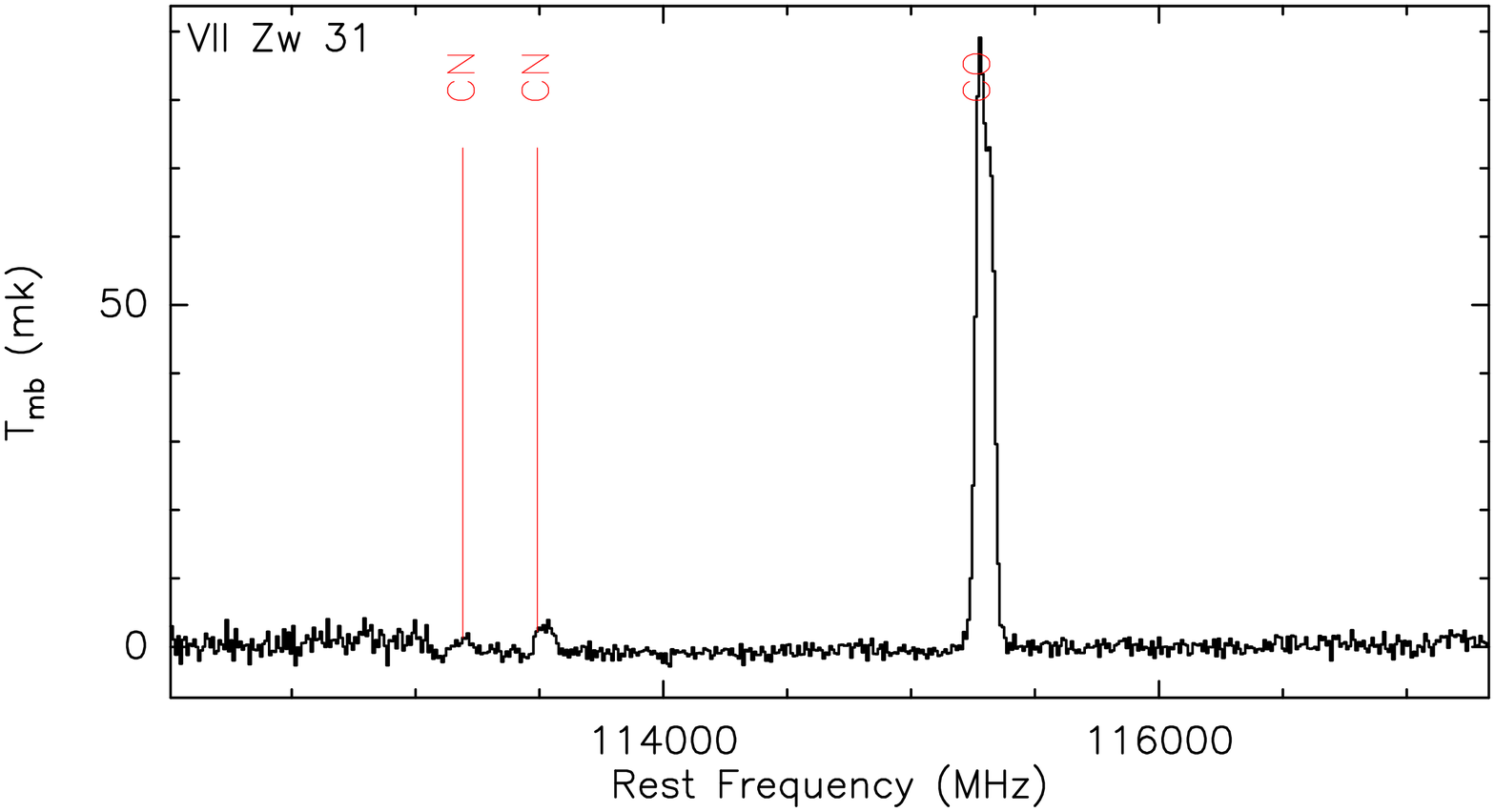}
\includegraphics[width=3.2in]{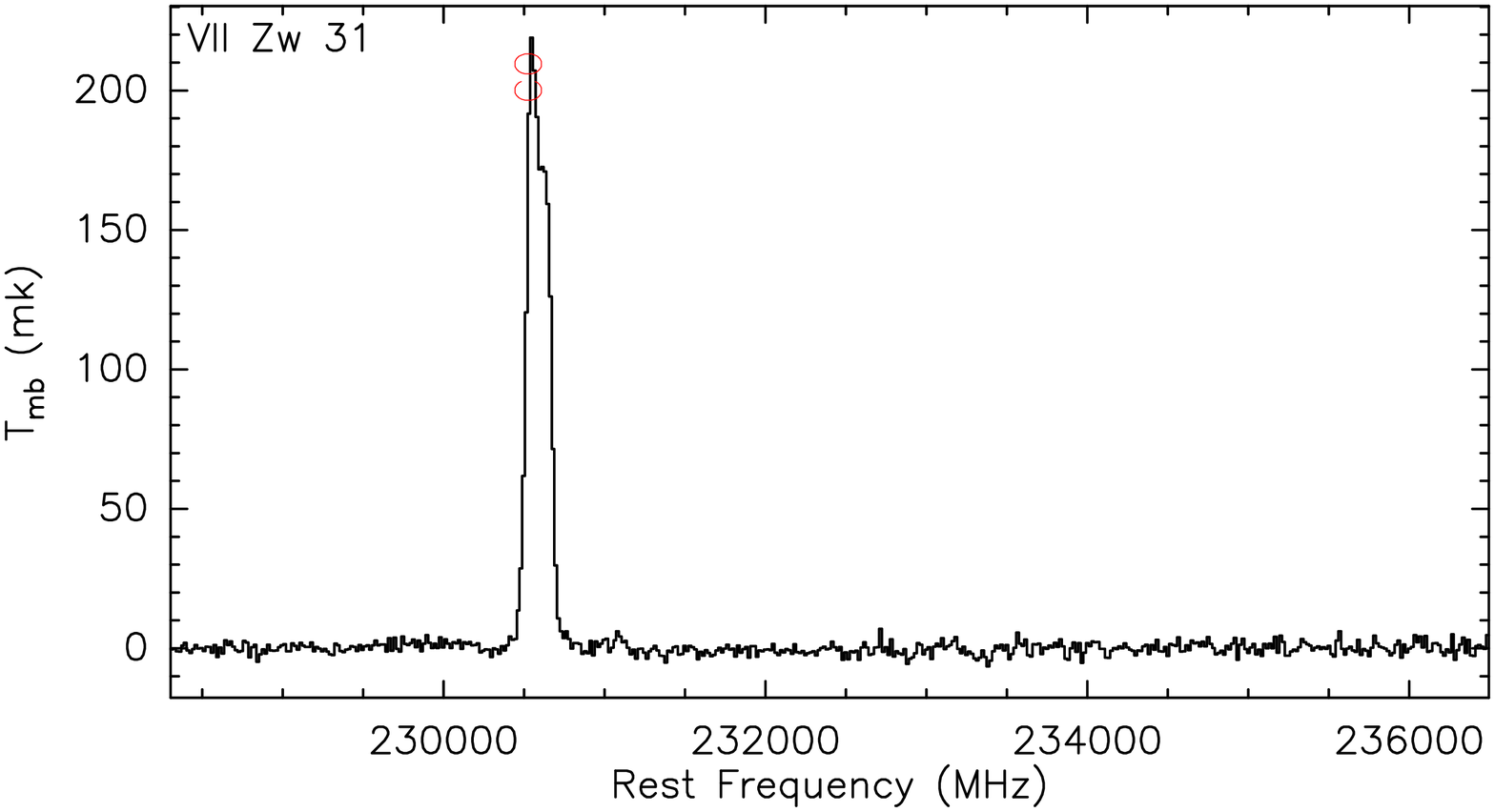}
\includegraphics[width=3.2in]{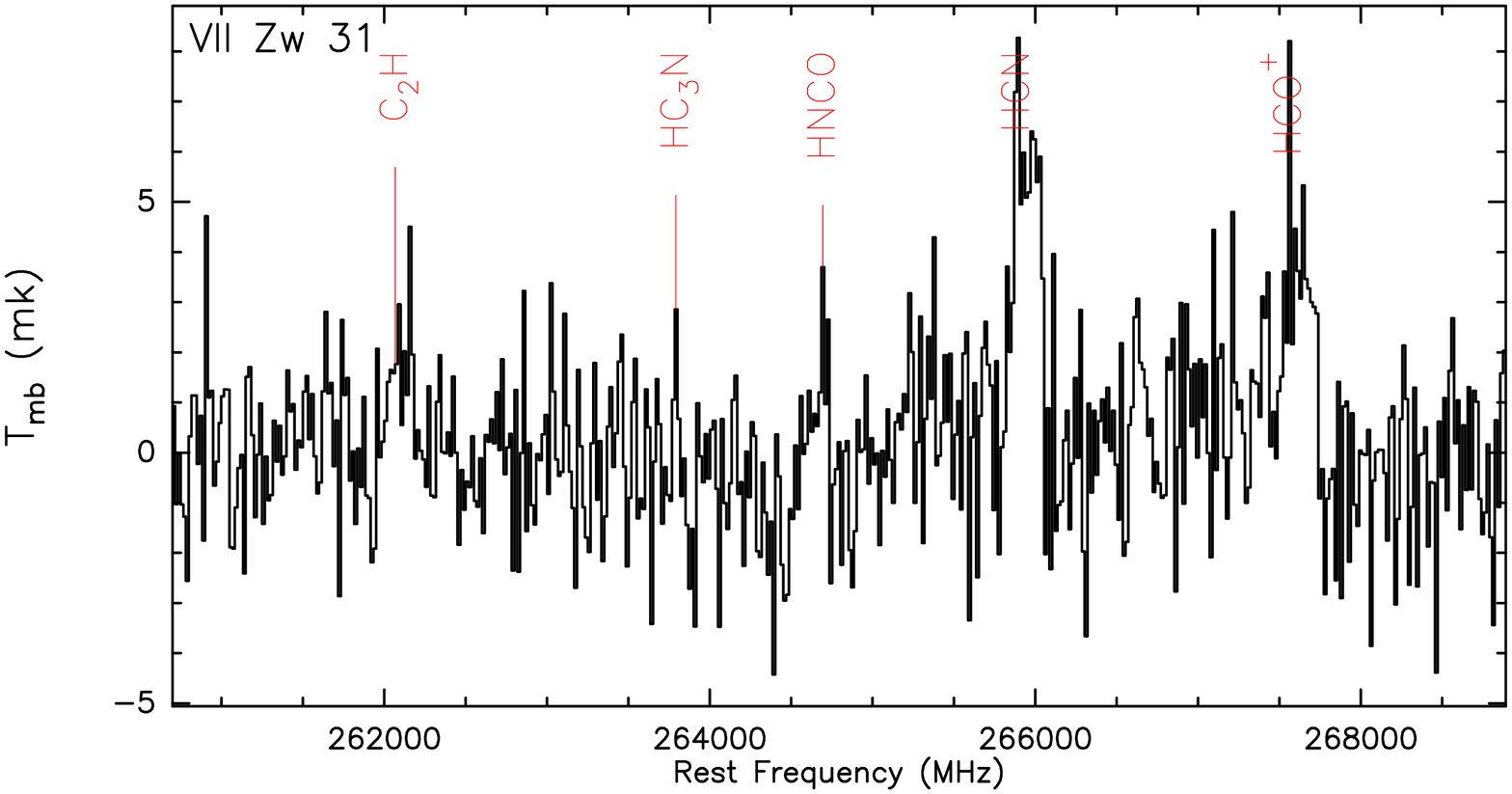}

\vspace*{-0.2 cm} \caption{Molecular species of the VII Zw 31. Left: band ranges from 108 GHz-111 GHz, with RMS of 1.15 mK at the  velocity resolution of -23.48 km\,s$^{-1}$.  
Middle: band ranges from 217 GHz-224 GHz, with   RMS  of 2.08 mK at the velocity resolution of  21.68 km\,s$^{-1}$.  
 Right: band ranges from 248 GHz-255 GHz, with RMS of 1.59 mK at the velocity resolution of  18.77 km$\cdot$s$^{-1}$. 
	\label{fig:VIIZW31}}
    \vskip-0.5pt
\end{figure*}

\bsp	
\label{lastpage}
\end{document}